\documentclass[%
 reprint,
amsmath,amssymb,superscriptaddress,nofootinbib,
 aps,
 prd
]{revtex4-2}
\usepackage{enumitem}

\usepackage[toc,page]{appendix}
\usepackage{booktabs}
\usepackage{lipsum}
\usepackage{bbm}
\usepackage{diagbox}
\usepackage{physics}
\usepackage{amsmath}
\usepackage{amssymb}
\usepackage{mathrsfs}
\usepackage[nolist,nohyperlinks]{acronym}
\usepackage[caption=false]{subfig}
\usepackage{url}
\usepackage{multirow}
\usepackage{xcolor}
\definecolor{dblue}{RGB}{25,25,125}
\usepackage[%
  colorlinks = true,
  urlcolor= dblue,
  linkcolor= dblue,
  citecolor= dblue
]{hyperref}
\usepackage{float}
\usepackage{graphicx}
\usepackage{capt-of}
\usepackage{tikz-cd}
\usepackage{tikz}
\usetikzlibrary{shapes,arrows}
\usetikzlibrary{shapes.geometric, arrows}
\usetikzlibrary{graphs}
\usetikzlibrary{cd}
\definecolor{dblue}{RGB}{25,25,125}
\tikzstyle{startstop} = [rectangle,rounded corners, minimum width=3cm,minimum height=1cm,text centered, draw=black]
\tikzstyle{io} = [trapezium, trapezium left angle = 70,trapezium right angle=110,minimum width=3cm,minimum height=1cm,text centered,draw=black]
\tikzstyle{process} = [rectangle,minimum width=2.5cm,minimum height=1cm,text centered,text width =2.5cm,draw=black]
\tikzstyle{decision} = [diamond,minimum width=2.5cm,minimum height=1cm,text centered,draw=black]
\tikzstyle{arrow} = [thick,->,>=stealth]
\usepackage{dcolumn}
\usepackage{bm}

\usepackage[commentmarkup=uwave]{changes}
\definechangesauthor[name=Max, color=magenta]{MI}
\definechangesauthor[name=Will, color=cyan]{WF}
\definechangesauthor[name=Haowen, color=red]{HZ}

\definecolor{mpl_red}{HTML}{D62728}

\newcommand{\red}[1]{{#1}}

\newcommand{\D}{\text{d}}

\newcommand{\mean}{\bm{\mu}}
\newcommand{\cov}{\bm{\Sigma}}
\newcommand{\data}{\bm{d}}

\begin{document}

\title{Multidimensional hierarchical tests of general relativity with gravitational waves}

\author{Haowen Zhong}
\email{zhong461@umn.edu}
\affiliation{
 School of Physics and Astronomy, University of Minnesota, Minneapolis, MN 55455, USA
}

\author{Maximiliano Isi}
\email{misi@flatironinstitute.org}
\affiliation{
Center for Computational Astrophysics, Flatiron Institute, 162 5th Ave, New York, NY 10010, USA
}

\author{Katerina Chatziioannou}
\email{kchatziioannou@caltech.edu}
\affiliation{Department of Physics, California Institute of Technology, Pasadena, California 91125, USA}
\affiliation{LIGO Laboratory, California Institute of Technology, Pasadena, California 91125, USA}

\author{Will M. Farr}
\email{wfarr@flatironinstitute.org}
\affiliation{
Center for Computational Astrophysics, Flatiron Institute, 162 5th Ave, New York, NY 10010, USA
}
\affiliation{
Department of Physics and Astronomy, Stony Brook University, Stony Brook, NY 11794, USA}

\date{\today}

\begin{abstract}
Tests of general relativity with gravitational waves typically introduce parameters for putative deviations and combine information from multiple events by characterizing the population distribution of these parameters through a hierarchical model.
Although many tests include multiple such parameters, hierarchical tests have so far been unable to accommodate this multidimensionality,
instead restricting to separate one-dimensional analyses and discarding information about parameter correlations.
In this paper, we extend population tests of general relativity to handle an arbitrary number of dimensions.
We demonstrate this framework on the two-dimensional inspiral-merger-ringdown consistency test, and derive new constraints from the latest LIGO-Virgo-KAGRA catalog, GWTC-3.
We obtain joint constraints for the two parameters introduced by the classic formulation of this test, revealing their correlation structure both at the individual-event and population levels.
We additionally propose a new four-dimensional formulation of the inspiral-merger-ringdown test that we show contains further information.
As in past work, we find the GW190814 event to be an outlier; the 4D analysis yields further insights on how the low mass and spin of this event biases the population results.
Without (with) this event, we find consistency with general relativity at the 60\% (92\%) credible level in the 2D formulation, and 76\% (80\%) for the 4D formulation.
This multi-dimensional framework can be immediately applied to other tests of general relativity in any number of dimensions, including the parametrized post-Einsteinian tests and ringdown tests.
\end{abstract}

\maketitle

\begin{acronym}
    \acro{GW}{gravitational wave}
    \acro{GR}{general relativity}
    \acro{CBC}{compact binary coalescence}
    \acro{BH}{black hole}
    \acro{BBH}{binary black hole}
    \acro{LVK}{LIGO-Virgo-KAGRA}
    \acro{PE}{parameter estimation}
    \acro{FAR}{false-alarm rate}
    \acro{GWOSC}{the Gravitational Wave Open Science Center}
    \acro{SSB}{solar system barycenter}
    \acro{SPA}{stationary phase approximation}
    \acro{PN}{post-Newtonian}
    \acro{BNS}{binary neutron star}
    \acro{CW}{continuous wave}
    \acro{IMR}{inspiral-merger-ringdown}
    \acro{NSF}{National Science Foundation}
    \acro{LKJ}{Lewandowsk-Kurowicka-Joe}
    \acro{GMM}{Gaussian mixture model}
    \acro{MCMC}{Markov chain Monte-Carlo}
    \acro{BIC}{Bayesian information criterion}
    \acro{ppE}{parametrized post-Einsteinian}
\end{acronym}

\section{Introduction}
\label{sec:intro}


In their first three observing runs (O1, O2, O3), LIGO~\cite{AdLIGO} and Virgo~\cite{AdVirgo},  
have detected \acp{GW} from over 90 \acp{CBC}~\cite{GWTC1,GWTC2,GWTC2.1, GWTC3}. 
These observations have not only opened up new frontiers for astrophysics and cosmology~\cite{GWTC1_BBH,GWTC1_Hubble,GWTC2_CBC,GWTC2_Gamma,GWTC2_Lensing,GWTC3_CBC,GWTC3_Cosmo,GWTC3_Gamma} but also bolstered support for \ac{GR}~\cite{GWTC1-TGR,GWTC2-TGR,GWTC3-TGR}.
Each individual \ac{GW} detection furnishes a test of \ac{GR}, leading our cumulative sensitivity to increase with the number of observations.
The expanding catalog of events calls for robust statistical methods to combine these tests and produce constraints on deviations from \ac{GR} from sets of detections~\cite{Zimmerman:2019wzo,Hier,Isi:2022cii,Pacilio:2023uef,Payne:2023kwj,Magee:2023muf,Essick:2023upv}.
\citet{Hier}, building upon work by~\citet{Zimmerman:2019wzo}, proposed a hierarchical-inference framework~\cite{James:1961,Lindley:1972,Efron:1977,Rubin:1981, Thrane:2018qnx, Vitale:2020aaz} for this purpose.
The framework enables a null-test of \ac{GR} that does not hinge on assumptions about the true theory of gravity or about how deviations manifest in different events.

Starting from  measurements of parameters controlling deviations away from \ac{GR} in individual events, the hierarchical framework characterizes the distribution of the true parameter across the population of events. 
Typically, parametrizations are constructed so that \ac{GR} is recovered in the limit of vanishing deviation parameters; this translates to a population distribution that is a delta function at the origin if \ac{GR} is correct, i.e., with the deviation vanishing for all events.
Since O2, population distributions have been obtained for \ac{ppE} deviations in the \ac{GW} phase coefficients~\cite{Yunes:2009ke,Li2012PN, agathos2014TIGER, Mehta:2022pcn}, ringdown analyses~\cite{Carullo:2019flw,Isi:2019aib}, and \ac{IMR} consistency tests~\cite{Ghosh2016qgn, Ghosh2017gfp}, among others~\cite{Hier,GWTC2-TGR,GWTC3-TGR,Ng:2023jjt}; recently, the framework has been extended to simultaneously model the \ac{GR} deviations and the astrophysical properties of sources~\cite{Payne:2023kwj}, as well as to factor in selection biases~\cite{Essick:2023upv,Magee:2023muf}.

However, so far, results have been limited to one-dimensional tests of \ac{GR}, which model a \emph{single} deviation parameter at a time, even for tests that are inherently multidimensional.
For example, ringdown tests introduce deviations in both the frequency and damping rate of one or multiple quasinormal modes, while the \ac{IMR} consistency test introduces two parameters that quantify agreement of the remnant mass and spin as inferred independently from high versus low frequencies of the signal.
In these cases, multidimensional posteriors are produced at the individual-event level, but then only the marginal distributions are considered when combining events.
On the other hand, the \ac{ppE} test is typically carried out by varying a single deviation coefficient at a time, in spite of the existence of multiple \ac{ppE} coefficients that should, in principle, be measured jointly (as has been done only occasionally~\cite{LIGOScientific:2016lio,Perkins:2022fhr,Pai:2012mv,Saleem:2021nsb}).

Previous work has considered how a deviation in one parameter can manifest in multiple coefficients when each of them is measured independently rather than  jointly, as is typically the case for the \ac{ppE} test.
In that case, the deviation is eventually detected by the one-dimensional hierarchical test of \ac{GR} given enough observations (Fig.~2 in Ref.~\cite{Hier}); however, there will be little indication regarding the true combinations of coefficients that can explain the observed departure from \ac{GR}, since individual-event measurements did not contain information about correlations across coefficients in the first place.
Furthermore, there have been no studies of the case in which such correlation information exists at the individual-event level but it is ignored at the catalog level, as has been the case for the \ac{IMR} and ringdown tests so far.
Reducing a multidimensional test to a single dimension discards information about potential correlations between the deviation parameters (both at the single-event and population levels) and decreases its sensitivity.

In this paper, we generalize the hierarchical test of \ac{GR} to handle an arbitrary number of deviation parameters simultaneously.
This allows us to properly deal with likelihood correlations at the individual-event level (induced by the measurement process, e.g., through parameter degeneracies), as well as potential correlation structure appearing in the intrinsic distribution of \ac{GR} deviations, were any of them to be detected.
Correlations in the intrinsic distribution of \ac{GR} deviations would be expected if the observed deviations were a function of binary parameters, like masses or spins---a common feature of several extensions to \ac{GR}.
We demonstrate an application in two and four dimensions on the \ac{IMR} test and GWTC-3 data.
The multidimensional analysis uncovers the structure of correlations between the test parameters, while confirming the data's consistency with GR with significance comparable to existing one-dimensional results.

The four-dimensional formulation also sheds further light on the role of the GW190814 event, an outlier for this test. 
With a remnant spin of $\chi_{\rm f} \approx 0.28$, this event is an outlier compared to the majority of the catalog that has $\chi_{\rm f} \approx 0.75$.
Since the remnant spin is correlated with the remnant spin inferred from pre-merger and post-merger data, ignoring the former leads to a preference for a nonzero value in the variance of the latter.
Such a variance would signal a GR deviation. 
The four-dimensional analysis gains access to this correlation and restores consistency with GR.

The organization of this paper is as follows. In Sec.~\ref{sec:method}, we detail the hierarchical formalism for an arbitrary-dimensional parameter space. 
In Sec.~\ref{sec:IMR} we summarize the two-dimensional \ac{IMR} test and introduce an extended four-dimensional formulation, which we argue can better encompass the structure of the data. 
Sections~\ref{sec:realdata} and~\ref{sec:realdata:4d} present results for a two- and four-dimensional test respectively, and discuss the role of GW190814 in both. 
We conclude in Sec.~\ref{sec:dis}.

\section{Method}\label{sec:method}

We adopt a hierarchical framework following~\citet{Hier}.
Consider $N$ events and $K$
beyond-GR parameters $\{\bm{\varphi}\}\equiv\{\varphi_1,\varphi_2,...,\varphi_K\}$.
Each individual event has a true underlying value $\{\widehat{\bm{\varphi}}\}$; GR is recovered for $\widehat{\bm{\varphi}}=0$.
We target the first two moments of the true distribution of $\{\widehat{\bm{\varphi}}\}$ by modeling it as a $K$-dimensional Gaussian $\mathcal{N}(\mean,\cov)$, where $\mean$ is a vector of length $K$ and $\Sigma$ is a $K\times K$ covariance matrix.
This is the $K$-dimensional generalization of the one-dimensional Gaussian of Refs.~\cite{Hier,GWTC2-TGR,GWTC3-TGR}.
The goal is to determine the posterior distribution of the $\mean$ and $\cov$ hyperparameters, which consist of $\frac{1}{2}K(K+3)$ numbers: $K$ components of $\mean$, and $\frac{1}{2}K(K+1)$ unique components of $\cov$, which is symmetric.
\ac{GR} is recovered in the limit that all means and variances (diagonal of $\cov$) vanish.

Disregarding selection effects~\cite{Essick:2023upv,Magee:2023muf}, the posterior distribution for $\mean$ and $\cov$ is
\begin{equation}
    p(\mean,\cov \mid \{\data_i\}_{i=1}^N)\propto p(\mean,\cov)\, \mathscr{L}(\{\data_i\}_{i=1}^N \mid \mean,\cov) \,,
    \label{eq:post}
\end{equation}
where $p(\mean,\cov)$ is the (hyper)prior, $\mathscr{L}(\{\data_i\}_{i=1}^N|\mean,\cov)$ is the hierarchical likelihood, and $\data_i$ is the data for the $i$th GW event; the constant of proportionality normalizes the distribution.
Selection effects can be accounted for by enhancing Eq.~\eqref{eq:post} with a detection efficiency factor following the usual procedure, e.g.,~\cite{Essick:2023upv,Magee:2023muf}.

\subsection{Hyperpriors}

We adopt separable (hyper)priors for $\mean$ and $\cov$: $p(\mean, \cov)=p(\mean)\, p(\cov)$.
For the mean vector $\mean=(\mu_1,...,\mu_{K})$, we choose an uncorrelated zero-mean Gaussian with some characteristic scale $\varsigma_{\mu,k}$ for each $k$, i.e.,
\begin{equation} \label{eq:prior_mu}
    p(\mean) = \prod_{k=1}^K \mathcal{N}(0, \varsigma_{\mu,k}^2)[\mu_k] \, .
\end{equation}
To avoid being overly restrictive, the prior scale $\varsigma_{\mu,k}$ should match or exceed the typical magnitude of the $\varphi_k$ measurements from individual events.\footnote{As an implementation detail, we usually rescale all our parameters by a (potentially dimensionful) constant before sampling, bringing all coefficients to unit scale and allowing us to set $\varsigma_{\mu,k} = 1$. This can be beneficial for non-affine samplers. We confirm that the prior does not affect subsequent results in App.~\ref{sec:sanity checks}.}
One could also replace this by a flat or Jeffreys prior, but Gaussian priors are computationally beneficial.
See \cite{Isi:2022cii} (including Appendix A therein) for a discussion of the number of events required for the likelihood to inform the posterior as a function of prior scale.

To set the prior $p(\cov)$ for the covariance matrix $\bm\Sigma$, we first decompose the matrix itself as
\begin{equation} \label{eq:prior_sigma}
    \cov=\bm\sigma^\intercal \mathscr{C}\bm\sigma \,,
\end{equation}
where the vector $\bm\sigma=(\sigma_0,...,\sigma_K)^\intercal$ encodes the intrinsic standard deviations of each parameter, and the matrix $\mathscr{C}_{kj}:=\Sigma_{kj}/\sqrt{\Sigma_{kk}\Sigma_{jj}}$ is the associated correlation matrix.
While $\bm\sigma$ encodes the typical magnitude of each $\widehat{\varphi}_k$, $\mathscr{C}_{ij}$ has unit-scale entries and reduces to the identity matrix if the parameters are uncorrelated; by construction, $\mathscr{C}$ is positive definite, with unit diagonal and $0 \leq |\mathscr{C}_{kj}| \leq 1$ for $k\neq j$.
We set priors for the scale vector $\bm\sigma$ and correlation matrix $\mathscr{C}$ separately, so that $p(\cov) = p(\bm\sigma)\,p(\mathscr{C})$.

For the scale vector $\bm\sigma$ prior, we choose an uncorrelated (truncated) normal distribution as we did for $\bm\mu$, but now with a set of scales $\varsigma_{\sigma,k}$. In other words, we set
\begin{equation}
    p(\bm\sigma) = \prod_{k=1}^K \mathcal{N}_{[0,\infty)}(0, \varsigma_{\sigma, k}^2)[\sigma_k]\, ,
\end{equation}
with the $[0, \infty)$ subscript indicating the additional constraint that $\sigma_k \geq 0$ for all $k$, and $p(\bm\sigma) = 0$ otherwise.
Here, again, the scale of the hyperprior $\varsigma_{\sigma, k}$ should match or exceed the expected scale of the $\widehat{\varphi}_k$'s.
As before, one could also replace this by a flat or Jeffreys prior.

For the correlation matrix, $\mathscr{C}$, we use the \ac{LKJ}~\cite{LKJ} distribution, which is a standard choice of prior for correlation matrices~\cite{Akinc_2018, Liue_2017, Tao_2022, Feng_2021}.
This is a probability density on the space of unit-diagonal, positive-definite correlation matrices; the density function can be defined as a power-law of the determinant, $\abs{\mathscr{C}}$, such that
\begin{equation} \label{eq:lkj}
    p(\mathscr{C}) = \mathrm{LKJCorr}(\mathscr{C}\mid\eta)\propto \abs{\mathscr{C}}^{\eta-1} \, ,
\end{equation}
for some shape parameter $\eta > 0$.
For any $\eta$, the \ac{LKJ} prior always has the identity matrix ($I$) as the expected value, i.e., $\mathrm{E}\left[ \mathscr{C}\right]_{\mathscr{C}\sim\mathrm{LKJ}} = I$, so that \emph{on average} there will be no correlations imposed across different $\widehat\varphi_k$'s.
On the other hand, the choice of $\eta$ controls the \emph{spread} of the distribution, and thus the amount of support for off-diagonal elements of $\mathscr{C}$, with larger values of $\eta$ more sharply favoring $\mathscr{C} = I$.

\begin{figure}[]
    \centering
    \includegraphics[width=0.45\textwidth]{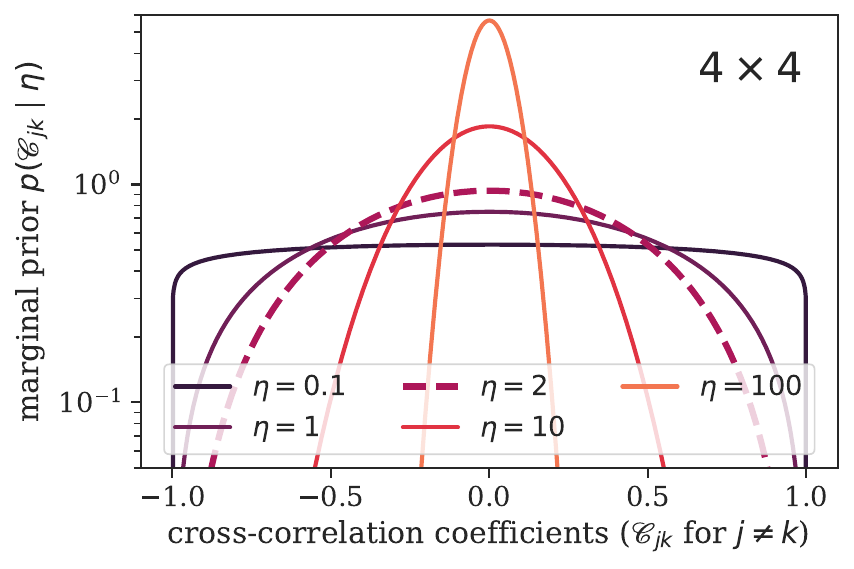}
    \caption{Marginal prior on the off-diagonal components of the correlation matrix $\mathscr{C}$ corresponding to the LKJ prior of Eq.~\eqref{eq:lkj}, assuming a $4\times 4$ correlation matrix, i.e., $K=4$ for different values of the shape parameter $\eta$. The marginal density follows a Beta distribution with shape parameters $\alpha=\beta = \eta -1 + K/2$, such that larger values of $\eta$ disfavor correlations more strongly.
    A dashed trace highlights our choice of $\eta = 2$ for the analyses in Secs.~\ref{sec:realdata} and \ref{sec:realdata:4d}.
    }
    \label{fig:lkj_marginal_prior}
\end{figure}

With this choice of prior on $\mathscr{C}$, each off-diagonal element $\mathscr{C}_{jk}$, $j\neq k$, will have a marginal prior given by a beta distribution, $B(\alpha,\beta)$, with shape parameters $\alpha=\beta = \eta - 1 +K/2$, for a $K\times K$ correlation matrix.
Concretely, if $\alpha=\beta=1$, then the density is uniform over correlation matrices; if $0 < \alpha=\beta <1$, the prior probability density drops for the identity matrix; if $\alpha = \beta > 1$, the prior peaks at the identity, with increasing sharpness for larger $\eta$.
For $K=4$, as corresponds to our case below, we display this marginal prior for different choices of $\eta$ in Fig.~\ref{fig:lkj_marginal_prior}, and representations of correlation matrices drawn from Eq.~\eqref{eq:lkj} in Fig.~\ref{fig:lkj_draws} in App.~\ref{appendixA}.

For concreteness, in our analyses below we choose a hyperprior $\eta=2$ that favors a weak correlation between beyond-GR parameters $\widehat{\bm\varphi}$ (dashed trace in Fig.~\ref{fig:lkj_marginal_prior});
this choice is not fixed and could be adjusted based on the specific problem at hand.
The full hyperprior is thus $p(\mean,\cov) = p(\mean)\, p(\bm\sigma)\, p(\mathscr{C})$, with factors given by Eqs.~\eqref{eq:prior_mu}, \eqref{eq:prior_sigma} and \eqref{eq:lkj}.

\subsection{Hierarchical likelihood}

The hierarchical likelihood, $\mathscr{L}(\{\data_i\} | \mean,\cov)$, is obtained from the likelihoods of individual events, $p(\{\data_i\} | \widehat{\bm{\varphi}})$, as
\begin{equation}
    \mathscr{L}(\{\data_i\}_{i=1}^N|\mean,\cov)=\int\D\widehat{\bm{\varphi}}\, p(\{\data_i\}_{i=1}^N \mid\widehat{\bm{\varphi}}) \, p(\widehat{\bm{\varphi}}|\mean,\cov) .
    \label{eq:hierlike}
\end{equation}
By construction of the population model, we have that $\widehat{\bm\varphi}\sim\mathcal{N}(\mean,\cov)$ so the second factor in the integrand is a Gaussian that can be evaluate in closed form, i.e., $p(\widehat{\bm{\varphi}}|\mean,\cov) = \mathcal{N}(\mean,\cov)[\widehat{\bm{\varphi}}]$.
The first factor is the likelihood of observing the data given true values of the deviation parameters $\widehat{\bm\varphi}$. Since each individual observation is independent, this separates into a product
\begin{equation}
p(\{\data_i\}_{i=1}^N \mid \widehat{\bm\varphi})=\prod_{i=1}^N p(\bm d_i \mid \widehat{\bm\varphi})\,.
\label{eq:like_prod}
\end{equation}
Each factor in this product is a $K$-dimensional likelihood obtained by applying the test of \ac{GR} to a single event in isolation.
Other parameters that may have been measured jointly with the $\widehat{\varphi}_k$'s have been implicitly marginalized over, assuming some fixed prior; it is often more appropriate to simultaneously model those parameters with the $\widehat{\varphi}_k$'s at the population level \cite{Payne:2023kwj}, as we revisit in Sec.~\ref{sec:imr:4d} below.

The individual-event likelihoods are typically estimated by reweighting posterior samples obtained under some fiducial sampling prior; then Eq.~\eqref{eq:hierlike} can be estimated via a Monte-Carlo sum~\cite{Vitale:2020aaz,Thrane:2018qnx}.
To further increase computational efficiency, we instead leverage the fact that the population model is a Gaussian and represent each individual-event likelihood through a \ac{GMM}. That is, we express the multidimensional single-event likelihood distribution for the $i$th event as a weighted sum of $N_{g,i}$ Gaussians, such that
\begin{equation}
p(\data_i \mid \widehat{\bm\varphi})\approx \sum_{j=1}^{N_{g,i}}w_j \, \mathcal{N}(\mean_i^{(j)},\bm{C}_i^{(j)})[\widehat{\bm\varphi}] \,.
\label{eq:GMM}
\end{equation}
Similar \ac{GMM} representations of individual event likelihoods have been used in the \ac{GW} literature before \cite[e.g.,][]{Golomb:2021tll}.
We can then analytically evaluate Eq.~\eqref{eq:hierlike} for each term in the \ac{GMM} and the hierarchical log-likelihood becomes
\begin{widetext}
\begin{equation}\label{eq:logL}
    \log\mathscr{L}(\{\data_i\}_{i=1}^N \mid \mean,\cov)=\sum_{i=1}^N\left[\log\Bigg(\sum_{j=1}^{N_{g,i}}w_j\frac{1}{\sqrt{(2\pi)^K \abs{\cov+\bm C_i^{(j)}}}}\exp\left[-\frac{1}{2}\left(\mean-\mean_i^{(j)}\right)^\intercal\left(\cov+\bm C_i^{(j)}\right)^{-1}\left(\mean-\mean_i^{(j)}\right)\right]\Bigg)\right] .
\end{equation}
\end{widetext}
We provide a detailed derivation of Eq.~\eqref{eq:logL} in App.~\ref{appendixA}.

Given this likelihood and the hyperprior discussed above, we sample the posterior for $\mean$ and $\cov$ per Eq.~\eqref{eq:post}.
The amount of support for \ac{GR} can be inferred by computing the probability for $\mean=\bm\sigma = 0$;
on the other hand, to the extent that there is support for $\bm\sigma >0$, the posterior for the correlation coefficients $\mathscr{C}_{jk}$ ($j \neq k$) will encode information about the nature of the deviation.

\subsection{Population-marginalized distribution}
\label{sec:ppd}

The result of the multidimensional hierarchical analysis is fully encompassed by the hyperposterior on $\mean$ and $\cov$.
Nevertheless, as is the case for the 1D case \cite{Hier,GWTC2-TGR}, it is sometimes useful to further compute a population-marginalized expectation for the deviation parameters, $\bm\varphi$.
This $K$-dimensional distribution, also known as the \textit{observed population predictive distribution}, represents our expectation for $\bm\varphi$ conditioned on the population properties inferred by the hierarchical analysis, marginalized over hyperparameters.
In arbitrary dimensions, this is formally
\begin{equation}
\label{eq:ppd}
    p(\bm\varphi \mid \left\{\data_i\right\}_{i=1}^N ) = \int \mathrm{d}\mean\, \mathrm{d} \cov\, p(\bm \varphi \mid \mean, \cov)\, p(\mean, \cov \mid  \left\{\data_i\right\}_{i=1}^N )\, ,
\end{equation}
and can be easily estimated by taking a draw from $\mathcal{N}(\mean, \cov)$ for each value of $(\mean,\cov)$ in the hyperposterior.
As in the 1D case, although convenient, this estimate has important limitations.
First, the shape of this distribution is directly related to the assumed Gaussian \textit{ansatz}, i.e., $\bm \varphi \sim \mathcal{N}(\mean, \cov)$, and therefore should not be taken to generally represent the shape of the true underlying distribution of deviation parameters.
Second, consistency with $\bm\varphi = 0$ is not a guarantee of consistency with \ac{GR}, as this can be satisfied even if $\sigma_k > 0$.
In spite of these limitations, the population expectation has been used to compare different catalog analyses in a succinct way~\cite{Hier,GWTC2-TGR,GWTC3-TGR}, so we demonstrate it below.

In the remainder of this paper, we apply this framework to the \ac{IMR} consistency test to demonstrate the advantages of the multidimensional hierarchical model in obtaining improved observational constraints.
We also validate our implementation with simulated data in App.~\ref{AppendixB} and further sanity checks in App.~\ref{sec:sanity checks}.


\section{Inspiral-merger-ringdown test}
\label{sec:IMR}

\newcommand{\Mpre}{M_\mathrm{f}^\mathrm{pre}}
\newcommand{\Mpost}{M_\mathrm{f}^\mathrm{post}}
\newcommand{\chipre}{\chi_\mathrm{f}^\mathrm{pre}}
\newcommand{\chipost}{\chi_\mathrm{f}^\mathrm{post}}
\newcommand{\dM}{\delta M_\mathrm{f}}
\newcommand{\dchi}{\delta \chi_\mathrm{f}}
\newcommand{\sM}{\mathscr{M}}
\newcommand{\schi}{\mathscr{X}}

\subsection{Traditional formulation}
\label{sec:imr:2d}

In this section, we provide an overview of the \ac{IMR} test~\cite{Ghosh:2016qgn, Ghosh:2017gfp}, which we will use to showcase our method.
The basic idea is to split the \ac{GW} data for each event into low- and high-frequency parts to obtain independent measurements of the source parameters, and then compare the two estimates for consistency.
The cut is performed in the Fourier domain to leverage the fact that the likelihood is diagonal in this space and so the two measurements are statistically independent, even though this does not technically separate the inspiral and merger-ringdown regimes exactly~\cite{Isi:2020tac,Cabero:2017avf}.
The cutoff frequency, $f_c^{\text{IMR}}$, is chosen for each event based on the merger frequency estimated from an analysis of the entire signal~\cite{GWTC3-TGR, Ghosh:2016qgn, Ghosh:2017gfp}.

Once a value of the cutoff has been chosen, the low ($f < f_c^{\text{IMR}}$) and high ($f > f_c^{\text{IMR}}$) frequency data are analyzed using a standard, Fourier-domain waveform model based in \ac{GR}, typically of the \textsc{IMRPhenom} family~\cite{Schmidt:2010it,Schmidt:2012rh,Hannam:2013oca,Khan:2019kot,Pratten:2020ceb,Garcia-Quiros:2020qpx,Pratten:2020fqn}.
The resulting posteriors are used to estimate the (detector-frame) remnant mass $M_\mathrm{f}$ and remnant spin $\chi_\mathrm{f}$ for each event; the estimates from low-frequency data are labeled $(\Mpre,\, \chipre)$, whereas those from high-frequency data are labeled $(\Mpost,\, \chipost)$.

If \ac{GR} is correct and the waveform is a good description of the data, we expect these two independent estimates of the remnant parameters to be in agreement.
We quantify departures from this expectation through the fractional deviations
\begin{subequations} \label{eq:def}
    \begin{equation}
        \dM \equiv 2\frac{\Mpre-\Mpost}{\Mpre + \Mpost}\,,
    \end{equation}
    \begin{equation}
        \dchi \equiv 2\frac{\chipre - \chipost}{\chipre + \chipost}\, ,
\end{equation}
\end{subequations}
so that \ac{GR} is recovered for $\dM = \dchi = 0$.
Since $\dM$ and $\dchi$ are not parameters we control in waveform models, we cannot directly obtain posterior estimates for these quantities.
Instead, their joint posterior is estimated by computing Eqs.~\eqref{eq:def} for independent draws from the $(\Mpre,\, \chipre)$ and $(\Mpost,\, \chipost)$ posterior ~\cite{GWTC1-TGR,GWTC3-TGR}; the likelihood on $(\dM, \dchi)$ is estimated by doing the same for the prior on $(\Mpre,\, \chipre)$ and $(\Mpost,\, \chipost)$, and then reweighting the posterior accordingly~\cite{GWTC2-TGR,GWTC3-TGR}.

The result of this process is an estimate of the two-dimensional likelihood function for $(\dM, \dchi)$ for each event.
Although these objects contain information about potential correlations between the two parameters, previous catalog analyses consider only one parameter at a time, i.e., they infer the population distribution of $\delta M_\mathrm{f}$ and $\delta \chi_\mathrm{f}$ separately~\cite{GWTC2-TGR,GWTC3-TGR}.
However, in doing so, they ignore potential correlations, with the drawbacks highlighted above.
To remedy this, we preserve the two-dimensional likelihood information for each event and apply our multidimensional hierarchical formalism, as encompassed by Eq.~\eqref{eq:post}.

\subsection{Extended formulation}
\label{sec:imr:4d}

The previous subsection describes the \ac{IMR} test as it has been formulated in the literature so far, yielding a two-dimensional parameter space $(\dM,\, \dchi)$.
However, we may go one step further by noting that, intrinsically, this is not a two-dimensional problem but a four-dimensional one: there are four basic quantities in this problem $(\Mpre,\, \chipre,\, \Mpost,\, \chipost)$, not two.
By considering only the fractional differences of Eq.~\eqref{eq:def}, we have disregarded half of the relevant parameters.

To take advantage of all the information inherent in the original test, we introduce two additional parameters, $\sM$ and $\schi$, defined as
\begin{subequations} \label{eq:def-extra}
    \begin{equation}
        \sM \equiv \frac{\Mpre+\Mpost}{2}\,,
    \end{equation}
    \begin{equation}
        \schi \equiv \frac{\chipre + \chipost}{2}\, .
\end{equation}
\end{subequations}
With this extension, the parameter space spanned by $\{\delta M_\mathrm{f}, \delta \chi_\mathrm{f}, \sM, \schi\}$ is equivalent to the initial parameter space spanned by $\{M_\mathrm{f}^{\mathrm{pre}},M_\mathrm{f}^{\mathrm{post}},\chi_\mathrm{f}^{\mathrm{pre}},\chi_\mathrm{f}^{\mathrm{post}}\}$ up to a coordinate transformation.

Restricting the hierarchical analysis to the $(\dM,\, \dchi)$ subspace would only be appropriate if these quantities were fully decoupled from $\sM$ and $\schi$ at the individual-event level, i.e., if the single-event likelihoods displayed no correlations across the two subspaces.
However, there is no reason \textit{a priori} to expect this to be the case, and indeed this is not the case for existing events, see Fig.~\ref{fig:likelihoods-4d}.
If any degree of correlation is present across the two subspaces, ignoring $\sM$ and $\schi$ is equivalent to marginalizing over these quantities by assuming a fixed prior distribution, c.f., Eq.~\eqref{eq:like_prod}.
This distribution is determined by the prior chosen for $(\Mpre,\, \chipre,\, \Mpost,\, \chipost)$ when projected onto this subspace, and is not physically meaningful or guaranteed to match the observed data.
As long as there are any correlations across $(\dM,\, \dchi)$ and $(\sM,\, \schi)$, this will bias the catalog test of \ac{GR}.

This situation is similar to that identified by~\citet{Payne:2023kwj}, who noted that parameters controlling deviations from \ac{GR} may couple to astrophysical parameters, like the \ac{BH} masses and spins.
The solution in that case, as well as here, is to simultaneously model all relevant degrees of freedom hierarchically at the population level.
In our case, this means that we not only model the two-dimensional $(\delta M_\mathrm{f},\delta \chi_\mathrm{f})$ subspace, but rather the full $(\delta M_\mathrm{f}, \delta \chi_\mathrm{f}, \sM, \schi)$ space, applying the framework in Sec.~\ref{sec:method} in four dimensions.%
\footnote{Future studies could consider alternative parametrizations for the $\sM$ population to match the observed structure of compact-binary masses, e.g., a power law plus a Gaussian peak \cite{GWTC3_CBC}.}
Consistency with \ac{GR} is still established for $\mu_k = \sigma_k = 0$ in the $(\dM,\, \dchi)$ subspace alone, after marginalizing over all other (nuisance) hyperparameters, including those controlling $\sM$ and $\schi$.

In the following, we present results for both the traditional (2D) and extended (4D) formulations of this test.

\section{Analysis of GWTC events}\label{sec:realdata}

\begin{table*}
\centering
\begin{tabular}{l|rrrr|rr|rr}
\toprule
\multirow{2}{*}{\textbf{Hyperparameter}} & \multicolumn{8}{c}{\textbf{Parameters considered in the analysis}} \\
\cmidrule{2-9}
& $\delta M_\mathrm{f}$ & $\delta\chi_\mathrm{f}$& $\delta M_\mathrm{f}^\star$ & $\delta\chi_\mathrm{f}^\star$ & $\{\delta M_\mathrm{f}, \delta\chi_\mathrm{f}\}$ & $\{\delta M_\mathrm{f}, \delta\chi_\mathrm{f}\}^\star$ & $\{\delta M_\mathrm{f},\delta\chi_\mathrm{f},\mathscr{M},\mathscr{X}\}$ & $\{\delta M_\mathrm{f},\delta\chi_\mathrm{f},\mathscr{M},\mathscr{X}\}^\star$ \\
\hline
\hline
$\mu_{\delta M_\mathrm{f}}$ &$0.04^{+0.08}_{-0.07}$ &--&$0.05^{+0.08}_{-0.07}$&--&$0.00^{+0.07}_{-0.07}$ &$0.02^{+0.07}_{-0.06}$ & $-0.02^{+0.06}_{-0.06}$& $-0.02^{+0.07}_{-0.06}$\\
$\mu_{\delta\chi_\mathrm{f}}$ &-- & $-0.04^{+0.11}_{-0.11}$ &--&$0.01^{+0.10}_{-0.10}$&$-0.09^{+0.10}_{-0.10}$ &$-0.02^{+0.08}_{-0.08}$ &$-0.11^{+0.09}_{-0.10}$ & $-0.07^{+0.08}_{-0.07}$\\
$\mu_\mathscr{M} / M_\odot$ & -- & -- & --&-- &--&--&$74.04^{+7.96}_{-7.87}$ & $77.67^{+7.21}_{-6.86}$\\
$\mu_\mathscr{X}$ & -- & -- &-- &-- &--&--&$0.75^{+0.04}_{-0.05}$ &$0.80^{+0.02}_{-0.03}$ \\
\hline
\hline
$\sigma_{\delta M_\mathrm{f}}$ & $0.04^{+0.09}_{-0.04}$ &-- &$0.04^{+0.09}_{-0.04}$&--&$0.05^{+0.09}_{-0.04}$ & $0.03^{+0.06}_{-0.03}$&$0.04^{+0.07}_{-0.03}$ & $0.02^{+0.05}_{-0.02}$\\
$\sigma_{\delta\chi_\mathrm{f}}$ & -- &$0.14^{+0.16}_{-0.12}$ &--&$0.06^{+0.11}_{-0.06}$&$0.14^{+0.12}_{-0.10}$ &$0.03^{+0.06}_{-0.03}$ &$0.13^{+0.11}_{-0.11}$&$0.02^{+0.05}_{-0.02}$ \\
$\sigma_\mathscr{M}/M_\odot$ & -- & -- &-- &-- &--&--&$18.59^{+7.27}_{-4.61}$ &$15.75^{+6.91}_{-4.25}$ \\
$\sigma_\mathscr{X}$ & -- & -- &--&--&-- &--&$0.09^{+0.06}_{-0.04}$&$0.02^{+0.03}_{-0.02}$ \\
\hline
\hline
$\rho_{\delta M_\mathrm{f}\delta\chi_\mathrm{f}}$ & --& -- &--&--&$0.43^{+0.49}_{-0.96}$ &$0.15^{+0.67}_{-0.82}$ & $0.25^{+0.53}_{-0.72}$&$0.15^{+0.58}_{-0.69}$ \\
$\rho_{\delta M_\mathrm{f}\mathscr{M}}$ & -- & -- &-- &-- &-- &--&$0.20^{+0.52}_{-0.68}$& $0.14^{+0.55}_{-0.66}$\\
$\rho_{\delta M_\mathrm{f}\mathscr{X}}$ & -- & -- &-- &-- & --& --&$0.07^{+0.60}_{-0.65}$&$0.02^{+0.62}_{-0.64}$\\
$\rho_{\delta\chi_\mathrm{f}\mathscr{M}}$ & -- & -- &-- &-- &-- & --&$0.46^{+0.34}_{-0.60}$&$0.11^{+0.57}_{-0.67}$\\
$\rho_{\delta\chi_\mathrm{f}\mathscr{X}}$ & -- & -- &-- &-- &-- & -- &$0.57^{+0.30}_{-0.67}$ & $0.00^{+0.63}_{-0.62}$ \\
\bottomrule
\end{tabular}
\caption{Medians and 90\% credible intervals for all hyperparameters and from all analyses. The first column indicates the hyperparameter, and the following columns shows their recovered values in each analysis. The superscript $\star$ indicates that we ignore GW190814 in that particular analysis. The 1D results we quote here are from our own reanalyses on the GWTC-3 events which are consistent to the results reported in Ref.~\cite{GWTC3-TGR}. The remaining columns are $2d$ and $4d$ results with and without GW190814 respectively.}
\label{table:1}
\end{table*}

Here we apply our method to the events analyzed in Ref.~\cite{GWTC3-TGR} to obtain higher-dimensional \ac{IMR}-test constraints on deviations from \ac{GR}.
Reference~\cite{GWTC3-TGR} considered 18 \ac{CBC} signals and combined them using a one-dimensional framework applied to $\dM$ and $\dchi$ separately.
That analysis found preference for a nonzero variance in the $\dchi$ population, i.e., low support for $\sigma = 0$ in Fig.~5 of Ref.~\cite{GWTC3-TGR}.
The spread in the $\dchi$ distribution was found to be driven by GW190814~\cite{GW190814}, which yields nonvanishing $\dM$ and $\dchi$ measurements with high credibility (possibly because of the lack of a sufficiently loud merger-ringdown~\cite{GWTC2-TGR}); removing this event from the set restored consistency with \ac{GR}~\cite{GWTC3-TGR}.

We revisit those results with a multi-dimensional analysis of both the traditional (2D) and extended (4D) formulations of the \ac{IMR} test outlined above, with and without the inclusion of GW190814.
We make use of prior and posterior samples for individual events made available by the \ac{LVK} collaborations~\cite{data}.
Our hyperprior is as described in Sec.~\ref{sec:method}, with $\eta = 2$ and $\varsigma_{\mu} = \varsigma_{\sigma} = 1$ for all parameters except $\sM$, for which we set $\varsigma_{\mu} = \varsigma_{\sigma} = 100 \, M_\odot$.
We summarize our results with medians and 90\% credible intervals for all hyperparameters from all analyses in Table~\ref{table:1}.

\begin{figure}[]
    \centering
    \includegraphics[width=\columnwidth]{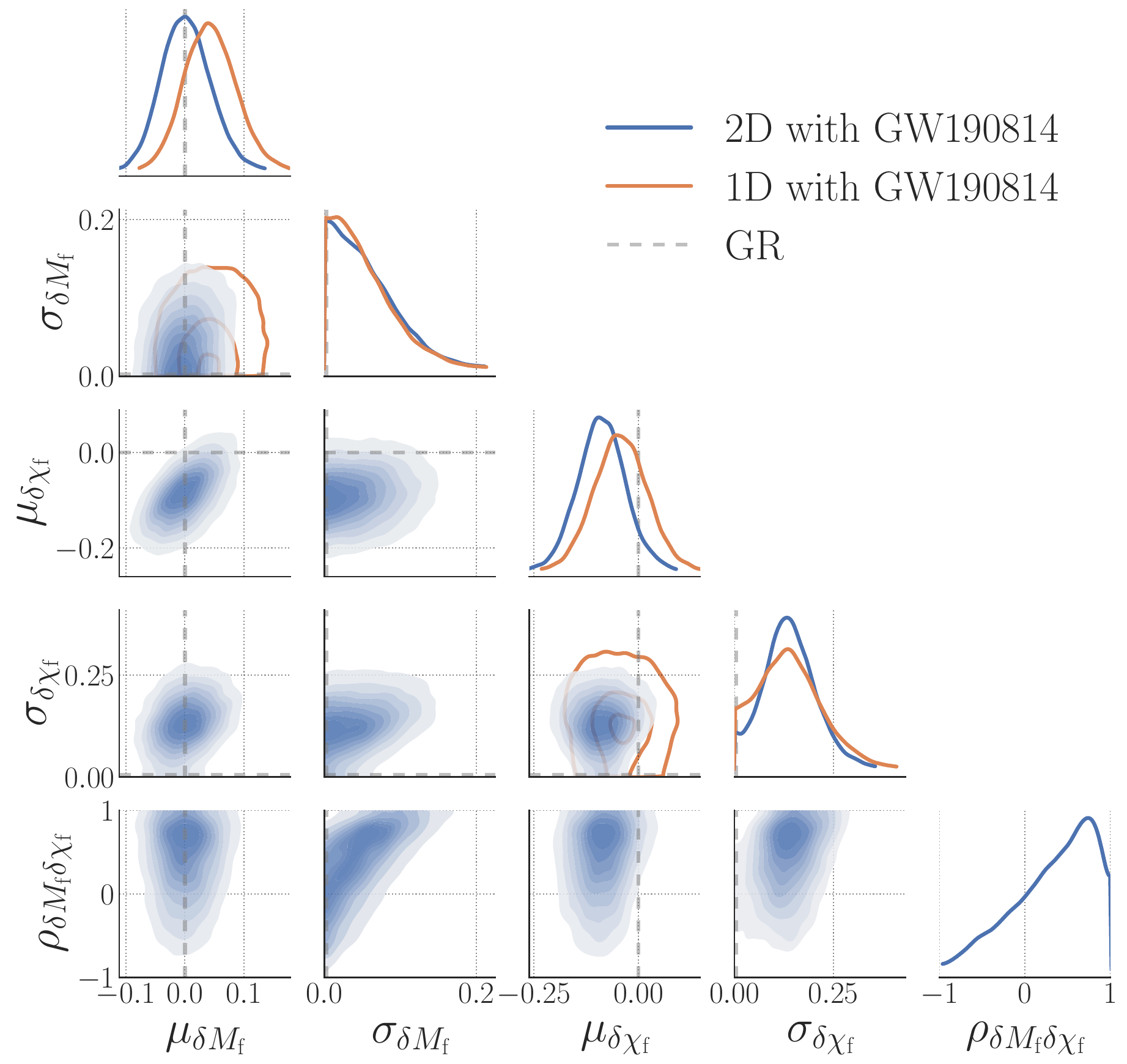}
    \caption{Result of the 2D hierarchical analysis on $\dM$ and $\dchi$ joint measurements (Sec.~\ref{sec:imr:2d}) from 18 GWTC-3 events including GW190814 (blue), compared to 1D analyses of the same events looking at $\dM$ and $\dchi$ separately (orange).
    The parameters are the mean and spread of $\dM$ ($\mu_{\dM}$, $\sigma_{\dM}$), the mean and spread of $\dchi$ ($\mu_{\dchi}$, $\sigma_{\dchi}$), and the correlation between the two ($\rho_{\dM\dchi}$); the 1D analyses only has access to marginals for $\dM$ and $\dchi$, so it can only measure their respective means and variances assuming no cross-correlation (orange sub-corners).
    Blue contours enclose probability mass at increments of 10\%, starting at 90\% for the outermost contour; orange contours enclose 90\%, 50\% and 10\% of the probability mass.
    \ac{GR} is recovered for $\mu_{\dM} = \sigma_{\dM} = \mu_{\dchi} = \sigma_{\dchi} = 0$ (gray dashed line).
    Inclusion of GW190814 leads to mild support for a deviation in $\dchi$.}
    \label{fig:2D}
\end{figure}

\subsection{Traditional 2D formulation}
\label{sec:realdata:2d}

We start by applying a two-dimensional version of the formalism to the traditional formulation of the test, in which only $\dM$ and $\dchi$ are explicitly considered (Sec.~\ref{sec:imr:2d}).
This analysis introduces five hyperparameters consisting of two population means ($\mu_{\dM}$, $\mu_{\dchi}$), two standard deviations ($\sigma_{\dM}$, $\sigma_{\dchi}$), and one correlation coefficient ($\rho_{\dM\dchi}$).
In the notation of Sec.~\ref{sec:method}, the mean vector is $\mean = (\mu_{\dM},\, \mu_{\dchi})$, the scale vector is $\bm\sigma = (\sigma_{\dM},\, \sigma_{\dchi})$, and the only off-diagonal component of the $2\times2$ correlation matrix is $\mathscr{C}_{12} = \mathscr{C}_{21} = \rho_{\dM\dchi}$.
Consistency with \ac{GR} is represented by $\mu_{\dM} = \mu_{\dchi} = \sigma_{\dM} = \sigma_{\dchi} = 0$, irrespective of $\rho_{\dM\dchi}$.

\subsubsection{Including GW190814}
\label{sec:realdata:2d:with}

We first show the result of the 2D analysis applied to all 18 events in our set, including GW190814.
Figure~\ref{fig:2D} shows posteriors for all four hyperparameters in the collective analysis (blue).
For comparison, we also display the result of 1D analyses that treat $\dM$ and $\dchi$ separately (orange), as was done in Ref.~\cite{GWTC3-TGR}.

As in Ref.~\cite{GWTC3-TGR}, we find that including GW190814 in our sample leads to mild support for a deviation from \ac{GR} through a nonvanishing $\sigma_{\dchi}$ (fourth diagonal panel).
This deviation is more apparent under the multidimensional formalism that models correlations between $\dM$ and $\dchi$, which also results in a preference for $\mu_{\dchi} < 0$ (third diagonal panel).
Accordingly, the 2D analysis recovers \ac{GR} at the \red{92\%} credible level, as opposed to \red{81\%} for the 1D $\dchi$ analysis (\red{64\%} for the 1D $\dM$ analysis).%
\footnote{A higher value for this credible level (quantile) corresponds to \emph{less} support for \ac{GR}, since it means that the posterior mass is distributed further away from the \ac{GR} point. We estimate it in practice as the fraction of posterior samples with probability density higher than the $\mu_{\dM} = \sigma_{\dM} = \mu_{\dchi} = \sigma_{\dchi} = 0$ point (marginalized over $\rho$); for the 1D analyses, this reduces to either the $\mu_{\dM} = \sigma_{\dM} = 0$ point or the $\mu_{\dchi} = \sigma_{\dchi} = 0$ point.}

The reason for the difference between the 2D and 1D analyses stems from the fact that there is evidence for correlations between the two deviation parameters, $\rho_{\dM\dchi} > 0$.
This is encoded in the structure of the 2D individual-event measurements: the 1D analyses, unable to access information contained in the 2D individual-event likelihoods, cannot distinguish such correlations from statistical uncertainty in either $\dM$ or $\dchi$, leading to broader hyper-posteriors and, correspondingly, degraded confidence in a deviation from \ac{GR} (fourth diagonal panel).
As a result of neglecting correlations, the 1D analyses also infers a potential offset from $\mu_{\dM}=0$ (top left panel).

On the other hand, the 2D analysis is able to infer that there are correlations between the $\dM$ and $\dchi$ measurements at the individual-event level and that, typically, some linear combination of the two parameters is better measured than each parameter alone.
 The 2D measurement can pin down both the $\dM$ and $\dchi$ quantities simultaneously, thus inferring that there are actually no clear anomalies in the $\dM$ distribution (top left panel), but that there are indeed anomalies in $\dchi$ (third and fourth diagonal panels).
 Furthermore, it also directly reveals that there are likely correlations between the two parameters (bottom right panel), and that this interaction is the dominant cause of variance in $\dM$ (bottom row, second column).
 This can be gleaned from the structure of the 2D likelihoods for individual events, and in particular the large negative value of $\dchi$ for GW190814, see Fig.~3 in Ref.~\cite{GWTC2-TGR}, or Fig.~\ref{fig:likelihoods-4d}.
 
All information about 2D correlations is destroyed when we first marginalize either quantity, as we do for the 1D $\dM$ or $\dchi$ analyses.
This highlights the power of the new method to better model deviations from \ac{GR} in multidimensional tests: when there is a departure from the null hypothesis (as is indeed the case here due to GW190814), the 2D analysis is not only better able to pick that up, but also sheds light on the nature of the putative deviation.

\subsubsection{Excluding GW190814}
\label{sec:realdata:2d:without}

\begin{figure}[]
    \centering
    \includegraphics[width=\columnwidth]{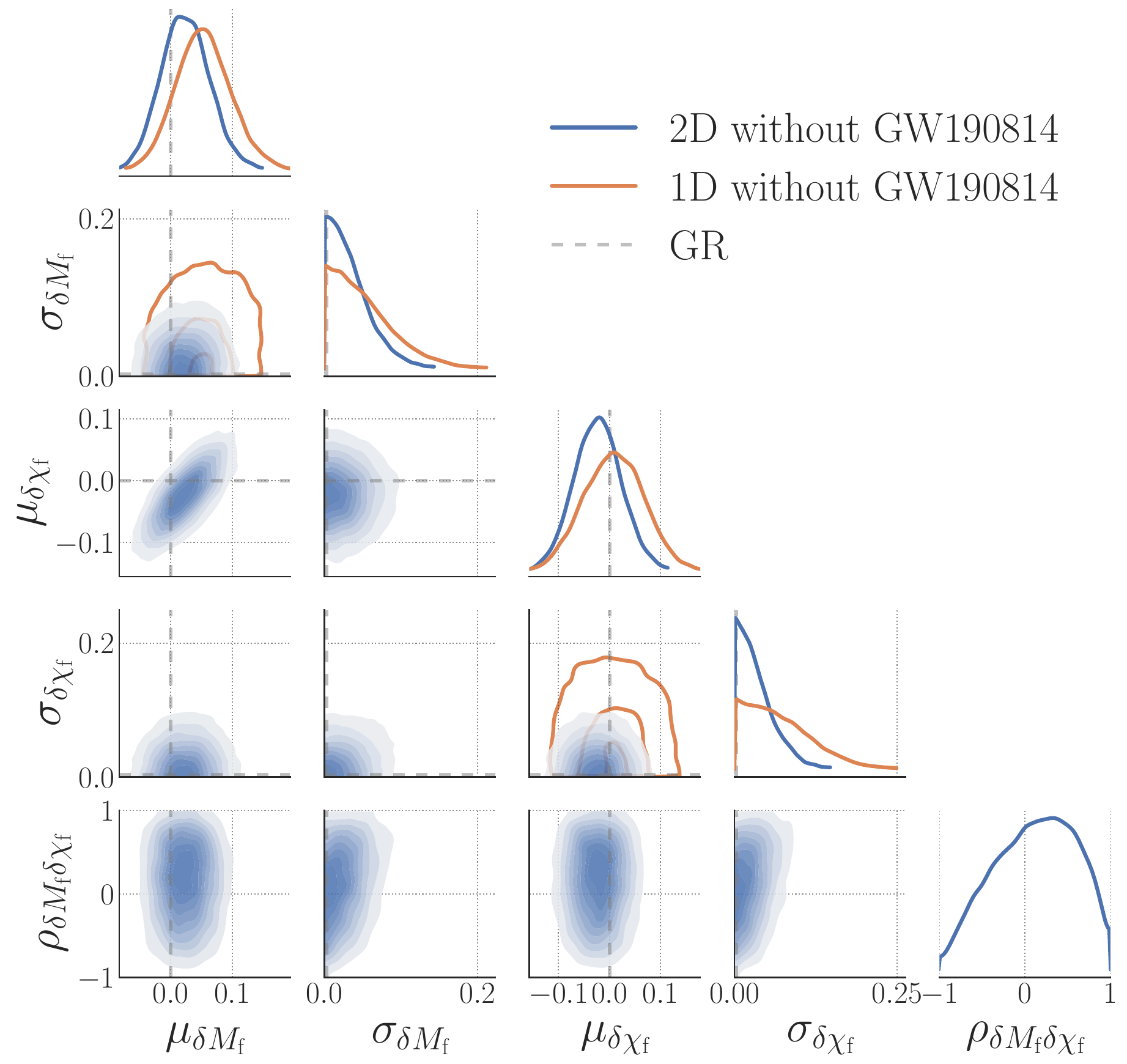}
    \caption{Same as in Fig.~\ref{fig:2D} but for an analysis that excludes GW190814. Exclusion of GW190814 removes the support seen in Fig.~\ref{fig:2D} for $\sigma_{\dchi} > 0$ and $\mu_{\dchi} < 0$.
    The 2D analysis (blue) is able to ascertain consistency with \ac{GR} with higher precision than the 1D analyses (orange).}
    \label{fig:2D_woGW190814}
\end{figure}

We repeat the analysis, but now excluding GW190814.
Figure~\ref{fig:2D_woGW190814} shows the results, again comparing the 2D framework (blue) to the traditional 1D framework (orange), as we did in Fig.~\ref{fig:2D}.
The exclusion of GW190814 has done away with what support there was for $\mu_{\dchi} <0$ or $\sigma_{\dchi} > 0$ in both the 2D and 1D analyses.
However, only the 2D analysis also displays reduced support for $\mu_{\dM} > 0$, as well as increased precision in the measurement of all parameters overall, i.e., tightening of blue versus orange distributions, as well as credible intervals in Table~\ref{table:1}. 
This leads to heightened credibility in \ac{GR}: the 2D analysis recovers \ac{GR} at the \red{60\%} credible level; the 1D analyses at \red{71\%} and \red{55\%} for $\dM$ and $\dchi$, respectively.

As discussed above, the 2D analysis is able to determine that the measurement process induces correlations in the joint $\dM$ and $\dchi$ likelihoods for individual events, so that some linear combination of the two parameters is typically better measured than either quantity alone.
This allows the 2D analysis to conclude that, in the absence of GW190814, the set of measurements is fully consistent with zero mean and no intrinsic spread in either deviation parameter ($\mu_{\dM} = \mu_{\dchi} = \sigma_{\dM} = \sigma_{\dchi} = 0$).
It does so with better precision than the 1D analyses because it can disentangle contributions to the observed variance in $\dM$ or $\dchi$ that are due to the correlations in the measurement process rather than an intrinsic spread in the population of true parameters.

Even in the absence of a deviation from GR, the correlation structure in the $\dM$ and $\dchi$ joint likelihoods leaves an imprint in the 2D hierarchical posterior, manifesting as a correlation in the inferred joint distribution for $\mu_{\dM}$ and $\mu_{\dchi}$ (second row, first column; also visible in Fig.~\ref{fig:2D}).
Irrespective of this correlation structure, there is no evidence for deviations from \ac{GR} in the set without GW190814, and $\mu_{\dM}=\mu_{\dchi}=0$ is well supported.
Since the posterior also favors vanishing variances ($\sigma_{\dM} = \sigma_{\dchi} = 0$), there are no strong preferences for positive or negative correlations and the posterior for $\rho_{\dM\dchi}$ resembles the prior (bottom right panel).

In summary, Fig.~\ref{fig:2D_woGW190814} again demonstrates the advantages of the 2D hierarchical framework, this time on a set of detections that show consistency with \ac{GR}.
Under these circumstances, the 2D analysis is able to achieve greater precision than the traditional 1D analysis by extracting information from the 2D likelihoods of individual events that is inaccessible to the 1D analyses.
Figure~\ref{fig:2D_woGW190814} also confirms that GW190814 is the cause for the deviations from \ac{GR} seen when analyzing the full GWTC-3 set, as was pointed out in Refs.~\cite{GWTC2-TGR,GWTC3-TGR}.

\subsection{Extended 4D formulation}
\label{sec:realdata:4d}

\begin{figure}[]
    \centering
\includegraphics[width=\columnwidth]{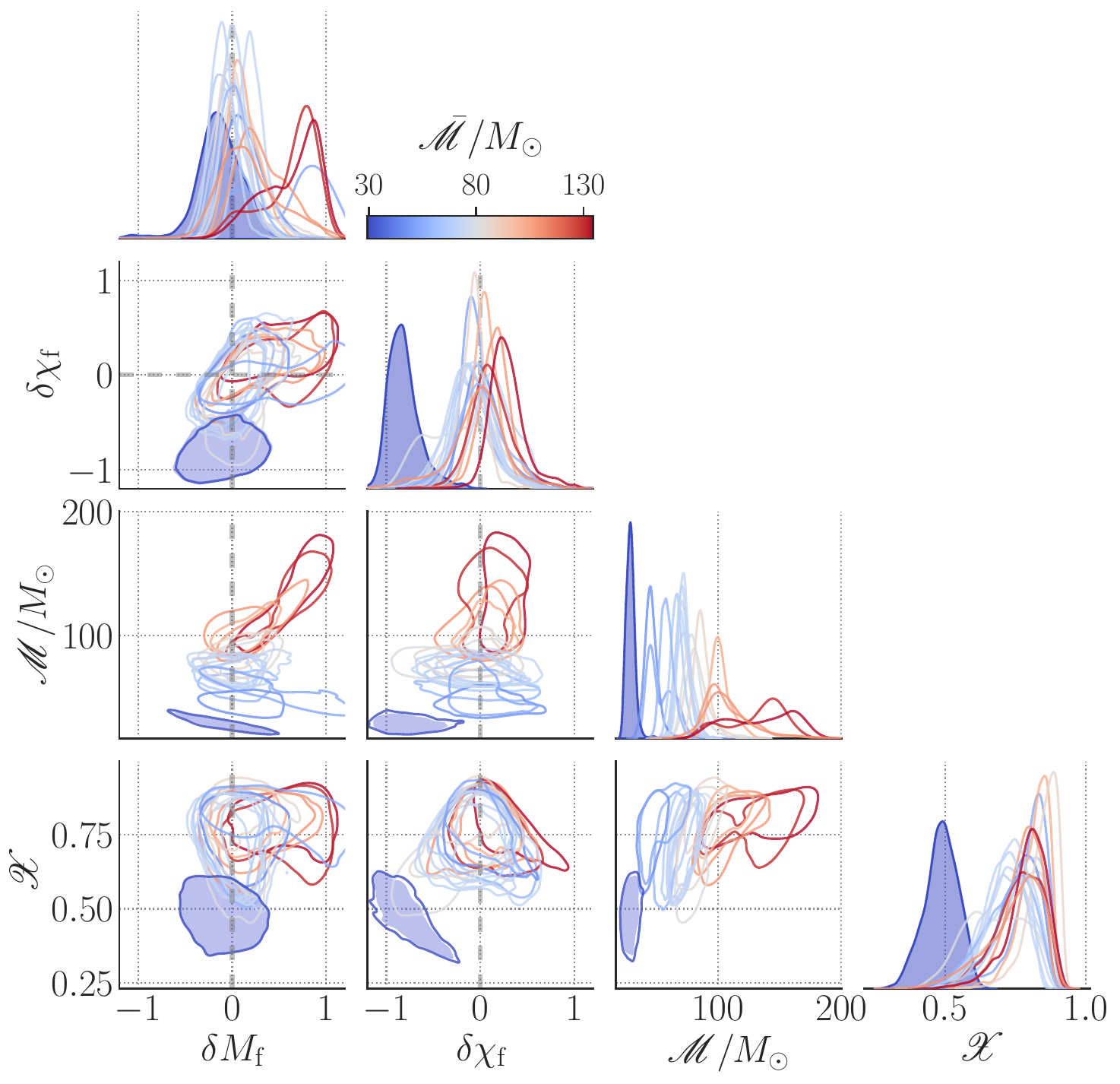}
    \caption{Four-dimensional likelihoods for each of the 18 events considered in the hierarchical IMR test analysis of GWTC data. The parameters correspond to the extended version of the IMR consistency test defined in Sec.~\ref{sec:imr:4d}. Contours enclose 90\% of the likelihood, colored by the inferred mean of the $\sM$ parameter (labeled $\bar{\sM}$), which is a proxy for the remnant mass; the null hypothesis requires $\dM = \dchi=0$ (dashed lines), irrespective of $\sM$ and $\schi$.
    The filled distribution highlights GW190814, an outlier in this population (Sec.~\ref{sec:realdata:4d}).}
    \label{fig:likelihoods-4d}
\end{figure}

We now turn to a hierarchical analysis of GWTC-3 over the full $4D$ parameter space comprised of $\{\dM,\, \dchi,\, \sM\, , \schi\}$, as described in Sec.~\ref{sec:imr:4d}.
Figure~\ref{fig:likelihoods-4d} shows the 4D individual-event likelihoods that make up the starting point for this analysis.
Whereas the 2D analysis marginalized over some \textit{ad hoc} implicit prior for the nuisance parameters $\sM$ and $\schi$, we here infer those populations simultaneously with $\dM$ and $\dchi$.
This analysis thus introduces 14 hyperparameters consisting of four population means ($\mu_{\dM}$, $\mu_{\dchi}$, $\mu_{\sM}$, $\mu_{\schi}$), four standard deviations ($\sigma_{\dM}$, $\sigma_{\dchi}$, $\sigma_{\sM}$, $\sigma_{\schi}$), and six correlation coefficients ($\rho_{\dM\dchi}$, $\rho_{\dM\sM}$, $\rho_{\dM\schi}$, $\rho_{\dchi\sM}$, $\rho_{\dchi\schi}$, $\rho_{\sM\schi}$).
As before, consistency with \ac{GR} is represented by $\mu_{\dM} = \mu_{\dchi} = \sigma_{\dM} = \sigma_{\dchi} = 0$, irrespective of the other parameters.

\subsubsection{Including GW190814}

\begin{figure*}
    \centering
    \includegraphics[width=\textwidth]{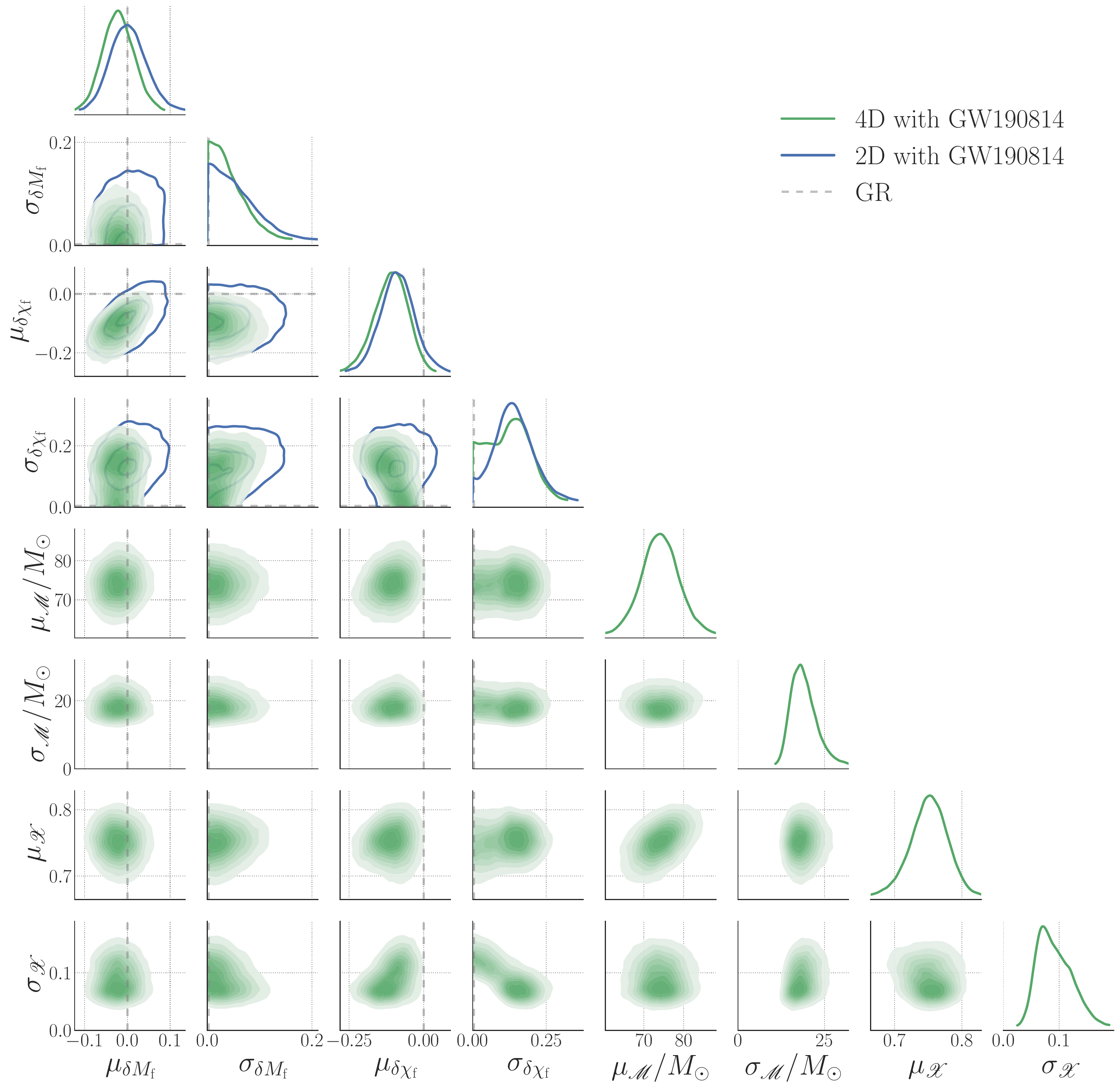}
    \caption{Subspace of the posterior measurement obtained from the 4D hierarchical analysis (Sec.~\ref{sec:imr:4d}) of 18 GWTC-3 events including GW190814 (green), compared to the 2D analysis of the same events from Fig.~\ref{fig:2D} (blue).
    The parameters are the mean and spread of $\dM$ ($\mu_{\dM}$, $\sigma_{\dM}$), the mean and spread of $\schi$ ($\mu_{\schi}$, $\sigma_{\schi}$), the mean and spread of $\dchi$ ($\mu_{\dchi}$, $\sigma_{\dchi}$), and the mean and spread of $\sM$ ($\mu_{\sM}$, $\sigma_{\sM}$); we omit posterior for the cross correlation parameters, which we show in Fig.~\ref{fig:14corner} of Appendix \ref{appendixC}.
    Green contours enclose probability mass at increments of 10\%, starting at 90\% for the outermost contour; blue contours enclose 90\%, 50\% and 10\% of the probability mass.
    \ac{GR} is recovered for $\mu_{\dM} = \sigma_{\dM} = \mu_{\dchi} = \sigma_{\dchi} = 0$ (gray dashed line).
    Inclusion of GW190814 leads to two largely distinct solutions per the 4D analysis: higher variance in $\schi$ and no variance in $\dchi$, or a lower variance in $\schi$ and a nonzero variance in $\dchi$---the latter of which corresponds to a deviation from \ac{GR} (or other systematic).}
    \label{fig:4D}
\end{figure*}

We begin with the full set of 18 GWTC-3 events, including GW190814, and analyze it under the 4D hierarchical framework.
Figure~\ref{fig:4D} displays a subspace of the posterior from the 4D analysis (green), excluding correlation coefficients for ease of display; the complete corner plot containing all 14 hyperparameters can be found in Fig.~\ref{fig:14corner} in App.~\ref{appendixC}.
In addition to the 4D result, Fig.~\ref{fig:4D} also shows the 2D result obtained in Fig.~\ref{fig:2D} for reference (blue).
Credible intervals for all hyperparameters can be found in Table.~\ref{table:1}.

The subspace of the IMR test corresponds to the upper left corner of Fig.~\ref{fig:4D}, showing the means and variances for the $\dM$ and $\dchi$ populations.
In relation to the 2D result, there is a slight reduction in overall variance, i.e., shrinkage of green contours relative to blue, and the 4D analysis recovers the null at a higher credible level of \red{80\%}, as opposed to \red{92\%} for the 2D analysis.
This suggests that there are correlations in the 4D likelihoods at the individual-event level, which was indeed the motivation for this extended analysis, see Sec.~\ref{sec:imr:4d} and Fig.~\ref{fig:likelihoods-4d}.

The existence of correlations across the $(\dM, \dchi)$ and $(\sM, \schi)$ subspaces is apparent in Fig.~\ref{fig:4D}.
In particular, the 4D analysis identifies a clear correlation between the variances of $\dchi$ and $\schi$, as can be seen from the $(\sigma_{\schi}, \sigma_{\dchi})$ panel in Fig.~\ref{fig:4D} (bottom row, fourth column).
Roughly speaking, there are two scenarios consistent with Fig.~\ref{fig:4D}: either (1) the $\schi$ population has standard deviation $\sigma_{\schi} \gtrsim 0.1$ and there is no variance in the $\dchi$ population; or (2) the $\schi$ population has standard deviation $\sigma_{\schi} \lesssim 0.1$ and there is a markedly nonzero variance in $\dchi$, which would imply a violation from \ac{GR} per this test.
The first of these scenarios also corresponds to a mean $\dchi$ closer to zero (bottom row, third column), and a likely lower variance in $\dM$ (bottom row, second column).

The structure in the 4D result helps further elucidate the anomalies in $\dchi$ in the 2D and 1D analyses when including GW190814 (Fig.~\ref{fig:2D}, as well as Refs.~\cite{GWTC2-TGR,GWTC3-TGR}).
Unable to directly access information about $\schi$, the 2D analyses effectively average over the possible scenarios outlined above, with some implied weighting imposed by the sampling prior on $\schi$ and $\sM$.
Accordingly, the result of the 2D analysis for $\sigma_{\dchi}$ does not correspond to either of the two modes exactly ($\sigma_{\dchi} = 0$ is disfavored but not excluded), although the second scenario appears to be upweighted.

We can understand the above observations by referring to the individual event likelihoods (Fig.~\ref{fig:likelihoods-4d}).
The measurement for GW190814 stands out in all dimensions, with the exception of $\dM$.
In particular, the structure of this likelihood shows clear correlations in the ($\schi, \dchi$) subspace: if $\schi$ were to take on a value closer to the bulk of the population (i.e., $\schi \approx 0.5$ or higher, closer to the population concentrated around $\schi \approx 0.75$), then we must have $\dchi \approx -1$; on the other hand, if $\dchi$ were to be closer to zero, then we must have that $\schi$ is much lower than the bulk of the population (i.e., $\schi \approx 0.35$).


\subsubsection{Excluding GW190814}

\begin{figure*}
    \centering
    \includegraphics[width=\textwidth]{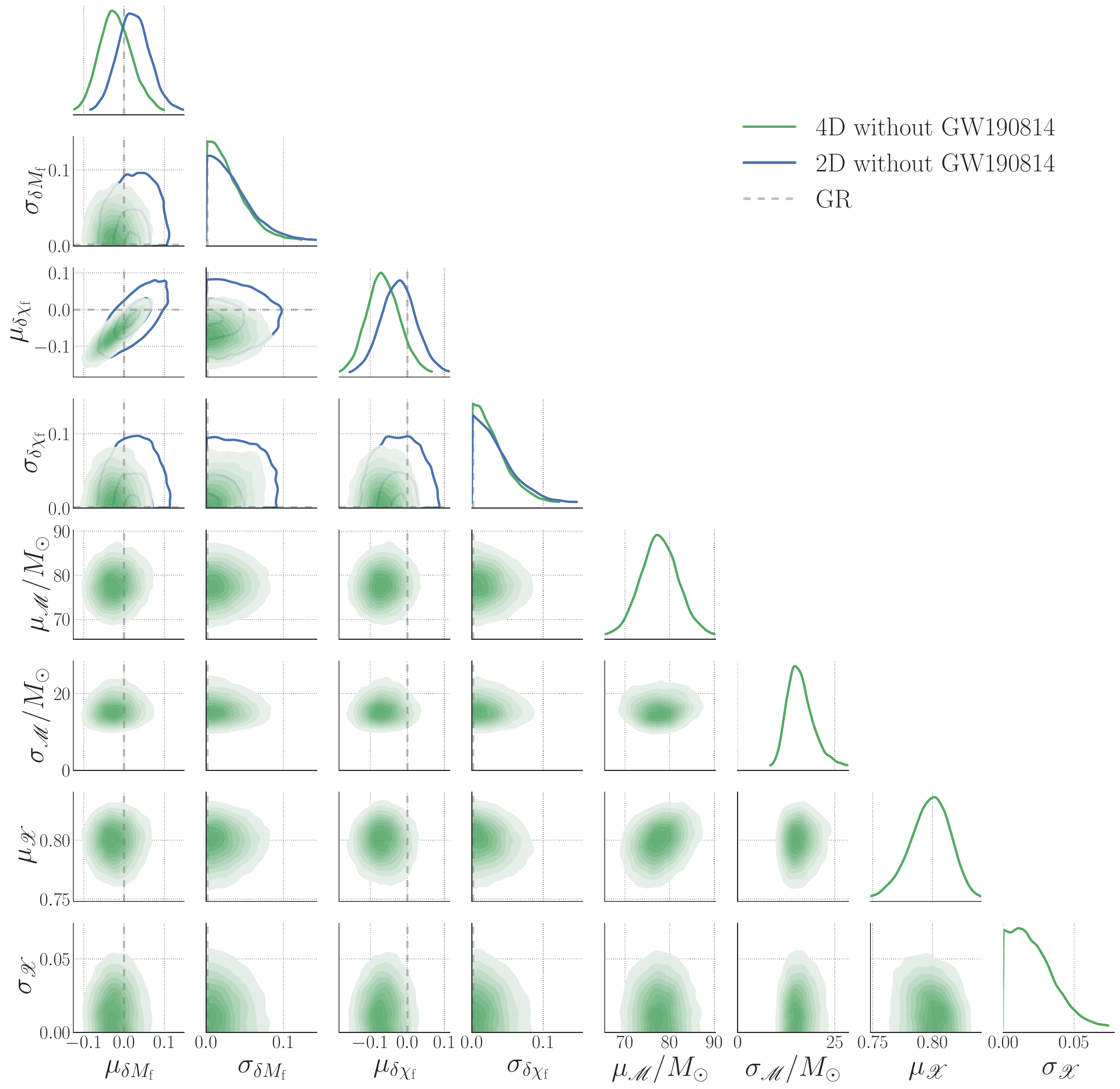}
    \caption{Same as Fig.~\ref{fig:4D} but now excluding GW190814. We show both a subspace of the posterior from the 4D hierarchical analysis (green) and the 2D result from Fig.~\ref{fig:2D_woGW190814} for comparison (blue); the corresponding full 14D corner plot including correlation parameters for the 4D hierarchical analysis is shown in Fig.~\ref{fig:14corner}.
    Excluding GW190814 gets rid of the bimodalities in the posterior from the 4D analysis.}
    \label{fig:4D_woGW190814}
\end{figure*}

We repeat the 4D hierarchical analysis, but now excluding GW190814 from our set of detections.
Figure~\ref{fig:4D_woGW190814} shows the result (green), in full analogy to Fig.~\ref{fig:4D}, but this time displaying the corresponding 2D analysis without GW190814 from Fig.~\ref{fig:2D_woGW190814} for reference (blue).
The full posterior including correlation coefficients for this analysis is shown in Fig.~\ref{fig:14corner} of Appendix \ref{appendixC}.

Without GW190814, the 4D hyperposterior is now unimodal, without outstanding interactions across the $(\dM, \dchi)$ and $(\sM, \schi)$ subspaces.
Now $\sigma_{\dchi} = 0$ is preferred (fourth diagonal panel) and, although the $\mu_{\dM}$ and $\mu_{\dchi}$ are slightly offset from zero (third row, first column), the overall posterior is broadly consistent with \ac{GR}, with the null hypothesis recovered at \red{76\%} credibility.
Further observations will be required to determine whether the slight shift in the means is simply due to statistical uncertainty or whether it represents a true systematic in the measurement or event selection process.

Although it does not directly factor into the test of \ac{GR}, this analysis infers low or vanishing variance in the $\schi$ parameter, with a typical mean value of $\mu_{\schi} \approx \red{0.8}$.
In terms of mass, the 17 events in this set are inferred to have a mean of $\mu_{\sM} \approx \red{80}\, M_\odot$, with a spread of $\sigma_{\sM} \approx \red{15} \, M_\odot$, see Table~\ref{table:1}.
With a remnant mass and spin of $M_{\rm f} \approx 25\, M_\odot$ and $\chi_{\rm f} \approx 0.28$ \cite{GW190814}, GW190814 would be a clear outlier for this population.

\subsection{Population-marginalized expectations}

\begin{figure*}
    \centering
    \includegraphics[width=0.8\columnwidth]{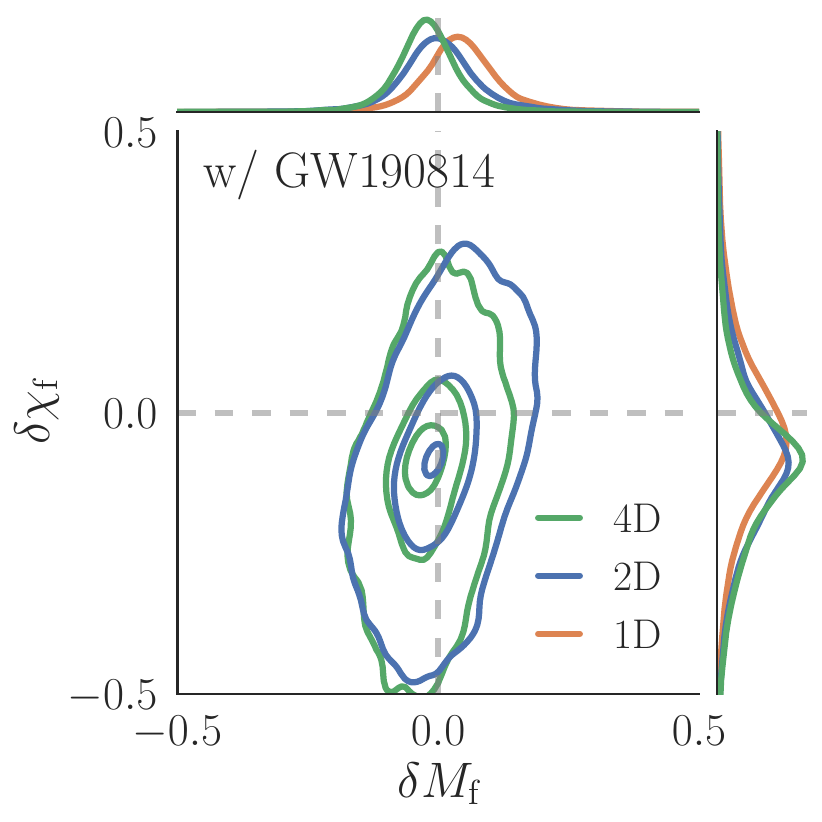}
    \qquad\includegraphics[width=0.8\columnwidth]{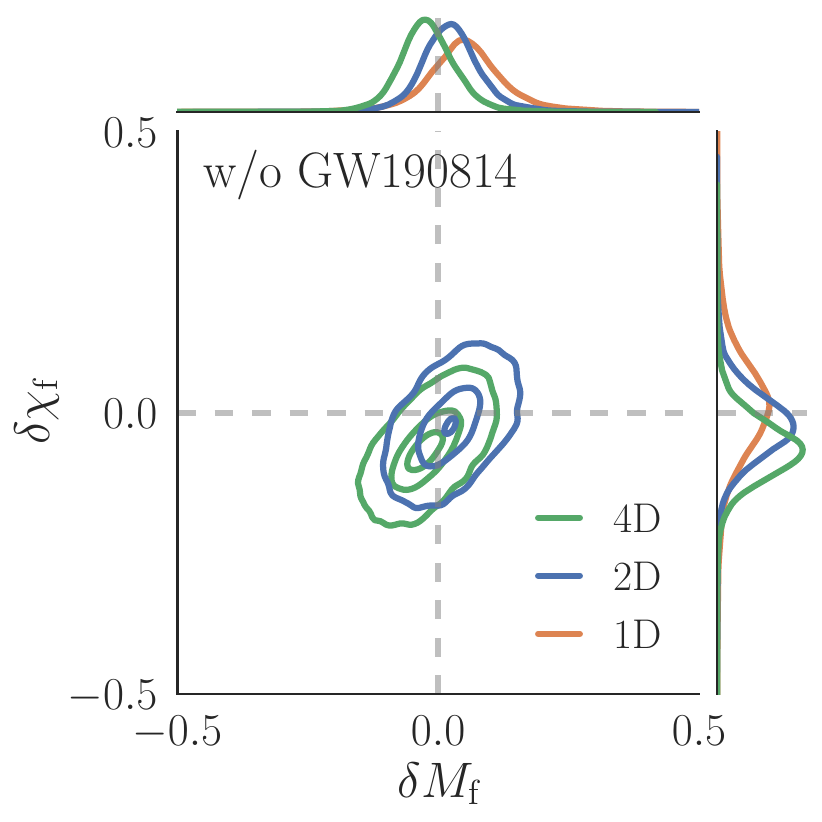}
    \caption{Population-marginalized distribution for the IMR test parameters $\dM$ and $\dchi$, Eq.~\eqref{eq:ppd}, as derived from all of the hierarchical analyses of GWTC-3 data presented in this paper: 4D (green), 2D (blue) and 1D (orange), with GW190814 (left) and without GW190814 (right).
    The results on the left correspond to the hyperposteriors in Figs.~\ref{fig:2D} and \ref{fig:4D}, while those on the right correspond to Figs.~\ref{fig:2D_woGW190814} and \ref{fig:4D_woGW190814}.
    Contours enclose 90\%, 50\% and 10\% of the probability mass; dashed lines mark the null expectation, $\dM=\dchi=0$.
    Excluding GW190814 leads to tighter population-marginalized distributions, especially for $\dchi$. See Table~\ref{tab:ppd} for constraints derived from these distributions.}
    \label{fig:ppd}
\end{figure*}

As described in Sec.~\ref{sec:ppd}, we may cast the hierarchical analysis result in a different light by computing the population-marginalized expectation (also known as the observed population predictive distribution) for $\dM$ and $\dchi$.
Although these derived distributions contain less information than the hyperposteriors, we compute them to facilitate comparison to past work and to derive constraints in directly in the $\dM$ and $\dchi$ space.
Figure~\ref{fig:ppd} shows the joint population expectation for $\dM$ and $\dchi$ derived from all the hierarchical analyses of GWTC-3 data presented in this section: 4D (green), 2D (blue) and 1D (orange), both with (left) and without (right) GW190814.
In all cases, these distributions are computed from a Monte-Carlo estimate, as described below Eq.~\eqref{eq:ppd}.

The population-marginalized distributions reveal some of the same features already described in the discussion of Figs.~\ref{fig:2D}--\ref{fig:4D_woGW190814}, but now directly in the space of $\dM$ and $\dchi$, rather than the hyperparameters.
The most obvious feature is the increased precision achieved by excluding GW190814 in the sample set, regardless of the dimensionality of the analysis, and particularly with regards to $\dchi$.
It is also notable that the population expectations derived from the multidimensional analyses (4D and 2D) carry information about the correlations in the inferred population means $\mu_{\dM}$ and $\mu_{\dchi}$, which here manifest as correlations between $\dM$ and $\dchi$ themselves.
These results also yield direct constraints on the $\dM$ and $\dchi$ values.
We report such constraints in Table~\ref{tab:ppd}, including for $\sM$ and $\schi$ which are not shown in Fig.~\ref{fig:ppd}.



\begin{table}[]
    \centering
    \caption{Population-marginalized constraints (median and 90\% credible symmetric interval).%
    \footnote{A star ($^\star$) denotes analyses \emph{excluding} GW190814.}}
    \label{tab:ppd}
    \begin{ruledtabular}
    \begin{tabular}{lrrrr}
    Analysis & $\delta M_{\rm f}$ & $\delta \chi_{\rm f}$ & $\mathscr{M}/M_\odot$ & $\mathscr{X}$ \\
    \midrule
1D & $0.04^{+0.14}_{-0.12}$ & $-0.04^{+0.28}_{-0.29}$ & - & - \\
1D$^\star$ & $0.05^{+0.13}_{-0.12}$ & $0.01^{+0.17}_{-0.17}$ & - & - \\
\midrule
2D & $0.00^{+0.14}_{-0.12}$ & $-0.09^{+0.28}_{-0.27}$ & - & - \\
2D$^\star$ & $0.02^{+0.10}_{-0.09}$ & $-0.02^{+0.11}_{-0.10}$ & - & - \\
\midrule
4D & $-0.02^{+0.11}_{-0.10}$ & $-0.10^{+0.23}_{-0.28}$ & $73.80^{+32.28}_{-32.75}$ & $0.75^{+0.16}_{-0.17}$ \\
4D$^\star$ & $-0.02^{+0.09}_{-0.09}$ & $-0.07^{+0.10}_{-0.09}$ & $77.78^{+28.12}_{-27.91}$ & $0.80^{+0.05}_{-0.05}$ \\
    \end{tabular}
    \end{ruledtabular}
\end{table}

\section{Conclusion}
\label{sec:dis}

In this paper, we have generalized the hierarchical inference framework for testing GR with gravitational wave observations from a single deviation parameter to an arbitrary number of parameters.
For tests that are formulated in terms of more than a single parameter, e.g., the ringdown and IMR tests, this generalization gains access to potential correlations between the test parameters both at the individual-event level, i.e., correlated likelihoods, and at the population level, i.e., correlated hyperparameters.

We applied the multi-dimensional framework to the \ac{IMR} consistency test using GWTC-3 events.
The IMR test divides a \ac{CBC} signal into high- and low-frequency portions and estimates the remnant mass and spin independently from each.
The test is parametrized via two deviation parameters, $\dM$ and $\dchi$, and two parameters that encode the remnant mass and spin, $\sM$ and $\schi$.
Formulations of the IMR test with reduced dimensionality, i.e., considering the population distribution of only $\dM$ and $\dchi$ separately, have previously yielded mild evidence for a violation of GR or other systematics (c.f., Fig.~\ref{fig:2D}), attributed to the GW190814 event.
Restoring the full four-dimensional formulation resolves this apparent deviation which is attributed to a correlation between $\schi$ and $\dchi$, c.f., Fig.~\ref{fig:likelihoods-4d}.
This application emphasizes the need to expand the dimensionality of tests of GR to all relevant parameters in order to avoid potential systematics from improper assumptions, such as ignoring correlations.

The expanded dimensionality means that more parameters need to be included in the analysis models and selection terms. 
This analysis focuses on multivariate Gaussian population distributions for all hyperparameters.
Although this model is reasonable for deviation parameters whose distribution cannot be motivated otherwise~\cite{Hier}, extended formulations could explore more complex distributions for the remnant mass and spin parameters. 
This situation is akin to the analysis of~\citet{Payne:2023kwj} that extended the parametrized phase deviation test to include the BH masses and spins and made use of distributions such as power-laws.
Moreover, the nature of the IMR test (that hinges on events with informative post- \emph{and} pre-merger data) makes estimating its selection effect particularly involved. 
We leave such extensions to future work with the expectation that their importance will increase as more events are detected and constraints are becoming more stringent.

\section*{Acknowledgements}

We thank Geraint Pratten for feedback on this manuscript.
The authors are grateful for computational resources provided by the LIGO Laboratory under NSF Grants PHY-0757058 and PHY-0823459. This material is based upon work supported by NSF's LIGO Laboratory which is a major facility fully funded by the National Science Foundation. H.Z. was supported by NSF Grant DGE-1922512.
The Flatiron Institute is funded by the Simons Foundation.
K.C.~was supported by NSF Grant PHY-2110111.
This paper carries LIGO document number LIGO-P2400214.

\appendix
\section{Hyperprior and likelihood derivation}\label{appendixA}

\begin{figure*}
    \centering
    \subfloat[$\eta = 0.1$]{\includegraphics[width=\textwidth]{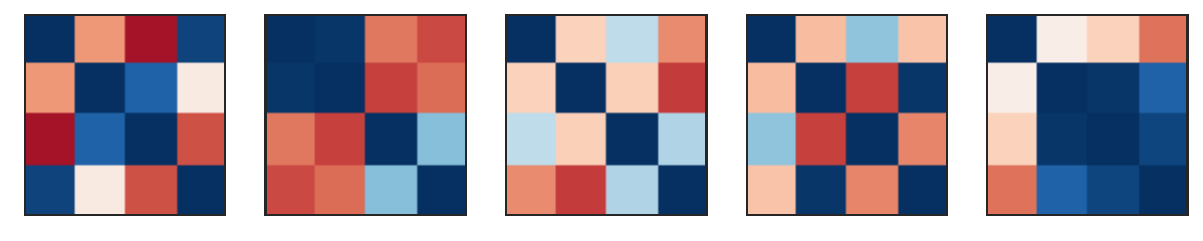}}\\
    \subfloat[$\eta = 1$]{\includegraphics[width=\textwidth]{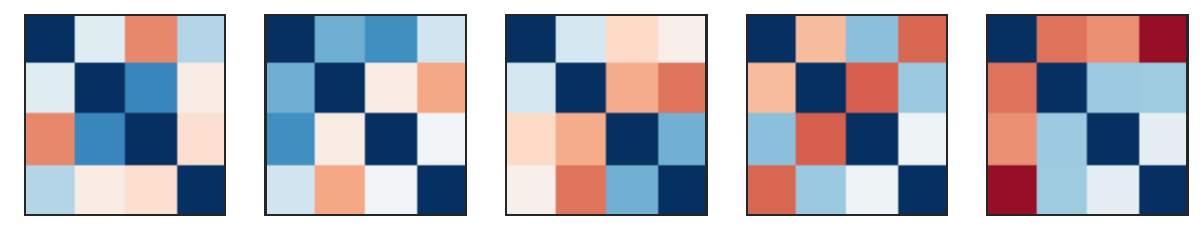}}\\
    \subfloat[$\eta = 2$]{\includegraphics[width=\textwidth]{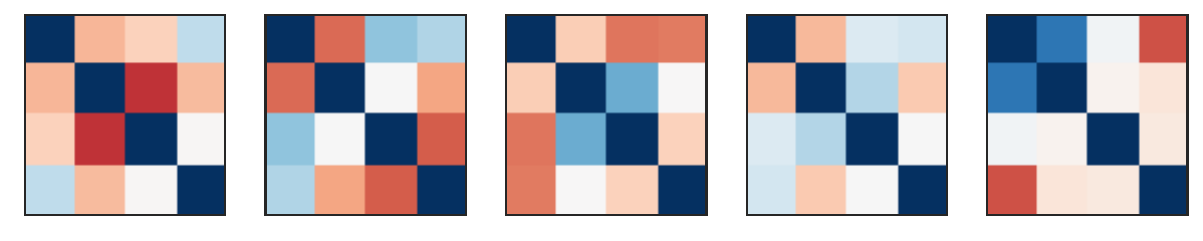}}\\
    \subfloat[$\eta = 10$]{\includegraphics[width=\textwidth]{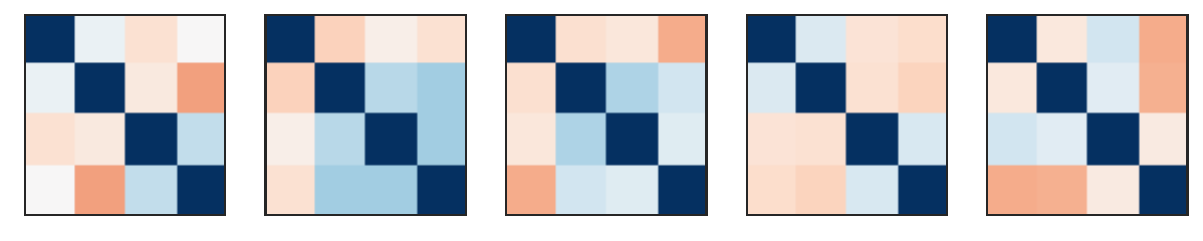}}\\
    \subfloat[$\eta = 100$]{\includegraphics[width=\textwidth]{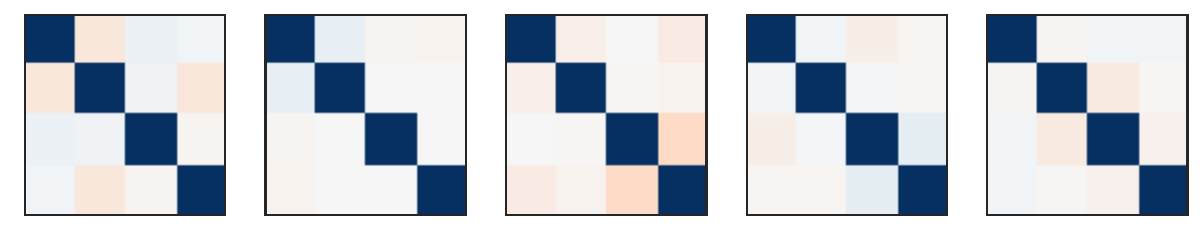}}\\
    \caption{Visualization of $4\times 4$ correlation matrices $\mathscr{C}$ drawn from an LKJ prior for different values of the shape parameter $\eta$ per Eq.~\eqref{eq:lkj}. For each $\eta$ (rows), we display five random draws (columns), with the color of each cell encoding the value of the corresponding $\mathscr{C}_{jk}$ entries: blues (reds) represents positive correlations $\mathscr{C}_{jk}>0$ ($\mathscr{C}_{jk} < 0$), and the intensity of the color encodes the magnitude of the correlation such that dark blue represents full positive correlation ($\mathscr{C}_{jk} = 1$), white represents no correlation ($\mathscr{C}_{jk}=0$), and dark red represents full anticorrelation ($\mathscr{C}_{jk}=-1$).
    The diagonal of $\mathscr{C}$ is always unity; the off-diagonal entries follow the marginal distributions shown in Fig.~\ref{fig:lkj_marginal_prior}.
    Larger values of $\eta$ favor the identity, i.e., lack of correlations, more strongly.}
    \label{fig:lkj_draws}
\end{figure*}

In this appendix, we provide a visualization of the covariance matric hyperprior in Fig.~\ref{fig:lkj_draws}, detail the derivation of Eq.~\eqref{eq:logL}, and provide further details on constructing the \ac{GMM} for each event.
We determine the number of Gaussians in the \ac{GMM}, $N_{g,i}$, and their parameters $\mean_i^{(j)}$ and $\bm{C}_i^{(j)}$ by optimizing the \ac{BIC}~\cite{Stoica_2004}
\begin{equation}
\text{BIC}:=k\ln(n)-2\ln(\hat{L})\,,
\end{equation}
where $k$ is the number of model parameters, $n$ is the sample size, and $\hat{L}$ is the maximized likelihood function value of the chosen model given the data. 
For each event, we employ a grid search~\cite{scikit-learn,sklearn_api}
to identify the optimal $N_{g,i}$ that minimizes the \ac{BIC}. Subsequently, we compute the best-fit $\mean_i^{(j)}$ and $\bm{C}_i^{(j)}$ 
for each chosen $N_{g,i}$. Typically, $N_g$ ranges from $\mathcal{O}(3-10)$.

To evaluate the integral in Eq.~\eqref{eq:hierlike}, we leverage the fact that the product of two Gaussians of arbitrary dimension can be refactored into the product of two different Gaussians as \cite{Bromiley:2013,Hogg:2020}
\begin{equation}
    \mathcal{N}(x|\mean_1,\cov_1)\, \mathcal{N}(x|\mean_2,\cov_2)=\mathcal{C}\mathcal{N}(x|\mean_3,\cov_3) \,,
\end{equation}
where $x$ denotes data samples, $\mean_i$ and $\cov_i~(i=1,2,3)$ are the mean vectors and covariance matrices of the corresponding multivariate Gaussians, and $\mathcal{C}$ is a normalization factor.
Explicitly, $\mathcal{C},\mean_3$ and $\cov_3$ are given by
\begin{subequations}
\begin{align} \label{eq:gaussian}
    \mathcal{C} &= \mathcal{N}(\mean_1 \mid \mean_2,\, \cov_1 + \cov_2)\, , \\
    \mean_3 &=(\cov_1^{-1}+\cov_2^{-1})^{-1}(\cov_1^{-1}\mean_1+\cov_2^{-1}\mean_2)\,,\\
    \cov_3 &=(\cov_1^{-1}+\cov_2^{-1})^{-1} \,,
\end{align}
\end{subequations}
where the Gaussian represented by $\mathcal{C}$ becomes a factor independent of the data $x$ and, in that sense, can be thought of as a normalizing constant.
With the help of Eq.~\eqref{eq:gaussian},
plugging Eq.~\eqref{eq:like_prod} and Eq.~\eqref{eq:GMM} back into Eq.~\eqref{eq:hierlike} yields Eq.~\eqref{eq:logL}.

\section{Simulated data}\label{AppendixB}

\begin{figure*}[]
    \centering
    \includegraphics[width=0.48\textwidth]{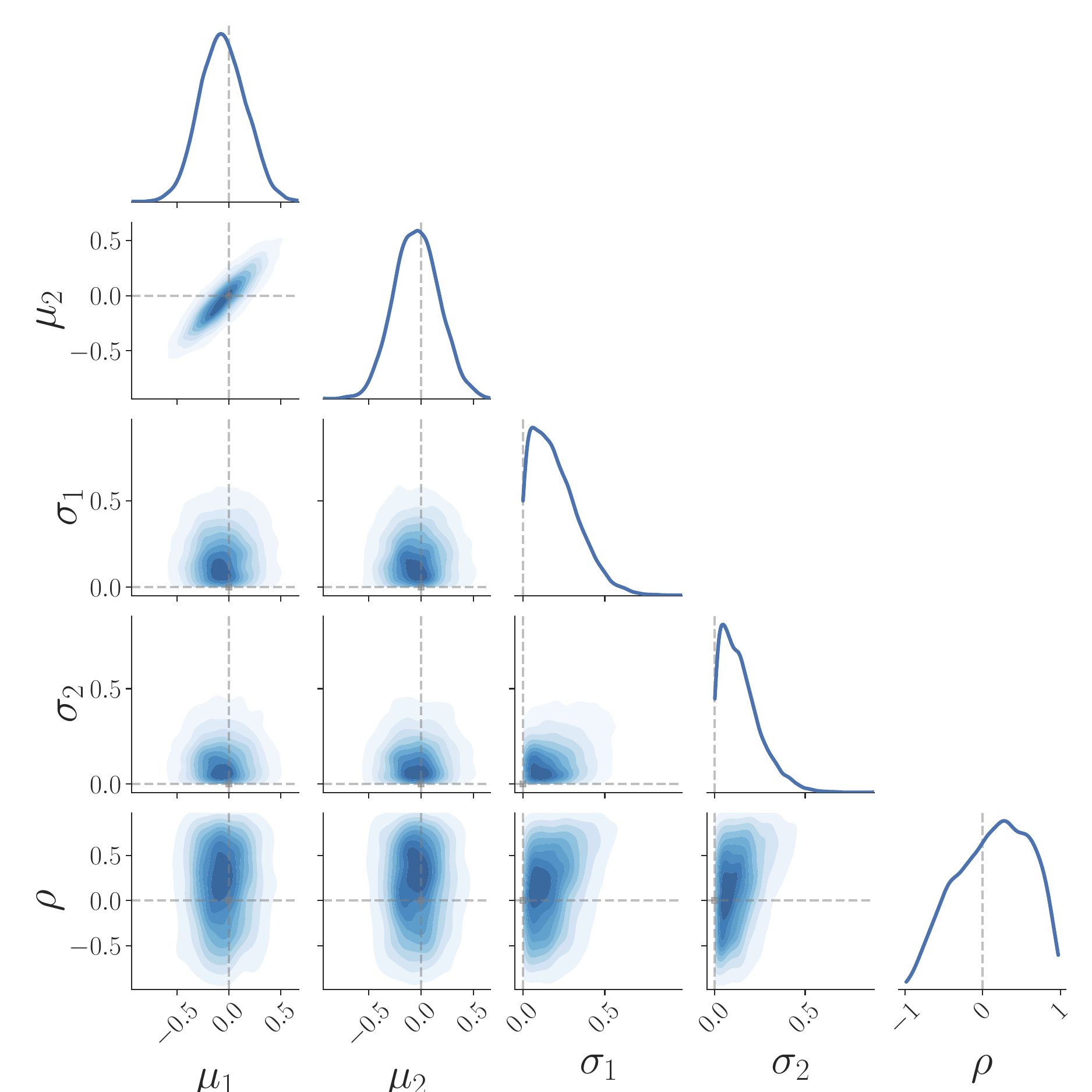}
    \includegraphics[width=0.48\textwidth]{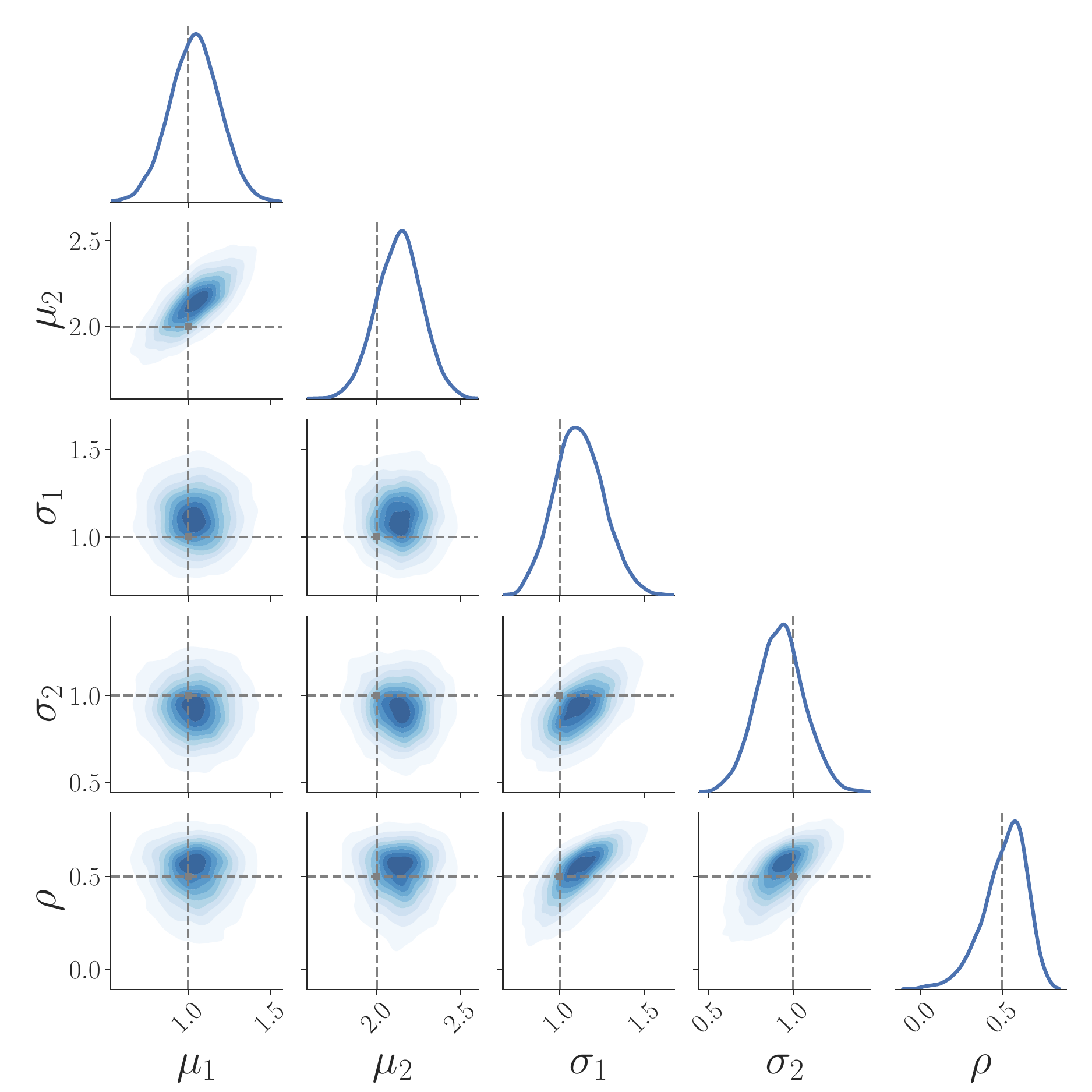}
    \caption{
    Marginalized posterior distributions on the hyperparameters $\mean$ and $\cov$
 of mock beyond-GR parameters $\varphi_1$ and $\varphi_2$ assuming GR is correct (left) or incorrect (right). The gray dashed lines indicate the location of true values. The correct parameters are always recovered within the distributions.}
    \label{fig:mock}
\end{figure*}

To verify the implementation of the multi-dimensional hierarchical analysis, we consider two simulated data scenarios: (a) GR is correct, and (b) GR is violated. The hyperparameters $\mean$ and $\cov$ are $\mean=(\mu_1,\mu_2)$ and
\begin{equation}
    \cov=\begin{pmatrix}
    \sigma_1^2&\rho\sigma_1\sigma_2\\
    \rho\sigma_1\sigma_2&\sigma_2^2
\end{pmatrix}\,.
\end{equation}
For the GR is correct case, $\widehat{\mean}=\bm0$ and $\widehat{\cov}=\bm 0$, while for the second case, we simulate deviations in both $\mean$ and $\cov$.
We simulate measurement uncertainty through a covariance matrix $\Sigma_\mathrm{obs}$, such that the simulated posterior samples, ${\varphi}_\mathrm{data}$, are drawn from
\begin{equation}
\left\{
\begin{aligned}
    &\widehat{\bm\varphi}\sim\mathcal{N}(\mean,\cov)\,,\\
    &\bm{\varphi}_\mathrm{obs}\sim\mathcal{N}(\widehat{\bm\varphi},\cov_\mathrm{obs})\,,\\
    &\bm{\varphi}_\mathrm{data}\sim\mathcal{N}(\bm\varphi_\mathrm{obs},\cov_\mathrm{obs})\,.
\end{aligned}
\right.
\end{equation}
To simulate the population, we set 
\begin{equation}\cov_\mathrm{obs}=
\begin{pmatrix}
        1&0.9\\
        0.9&1
\end{pmatrix}\,.
\end{equation} 
which corresponds to two strongly-correlated beyond-GR parameters.

Results are show in Fig.~\ref{fig:mock}.
For the case where GR is correct (left), we simulate 20 events and 1000 likelihood samples for each event. The true values of $\mean$ and $\cov$ are recovered at the 90\% credible level.
For the case where GR is incorrect (right), we simulate $\mean$ and $\cov$ as follows:
\begin{equation}
\widehat{\mean}=(1,2),\quad \widehat{\cov}=
\begin{pmatrix}
1&0.5\\
0.5&1
\end{pmatrix}\,.
\end{equation}
Compared to the case where GR is correct where 
$\widehat{\cov}$ and $\cov_\mathrm{obs}$ differ significantly, $\widehat{\cov}$ is now comparable to $\cov_\mathrm{obs}$. We simulate 100 events and 1000 likelihood samples for each and again recover the true parameters to within the 90\% credible level.

\section{Analysis sanity checks}\label{sec:sanity checks}

In this appendix, we confirm that (a) the choice of hyperprior $\eta$ does not affect the hyperposteriors, and (b) we recover the 1D results from the 2D analysis in the appropriate limit, i.e., for $\rho\to 0$ and no correlation between the individual-event $\delta M_\mathrm{f}$ and $\delta\chi_\mathrm{f}$ likelihoods. 
For the latter, we start with individual-event 2D samples denoted as $\{(\delta M_{\mathrm{f},i},\delta\chi_{\mathrm{f},i})\}_{i=1}^{N{s}}$. 
We then independently shuffle the sets $\{\delta M_{\mathrm{f},i}\}_{i=1}^{N{s}}$ and $\{\delta \chi_{\mathrm{f},i}\}_{i=1}^{N{s}}$ and create a new set of paired samples, $\{(\delta M_{\mathrm{f},{i'}},\delta\chi_{\mathrm{f},{i'}})\}_{i'=1}^{N{s}}$. This process removes any correlations between these two parameters in the individual-event likelihood.

\begin{figure*}
    \centering
    \includegraphics[width=1.0\textwidth]{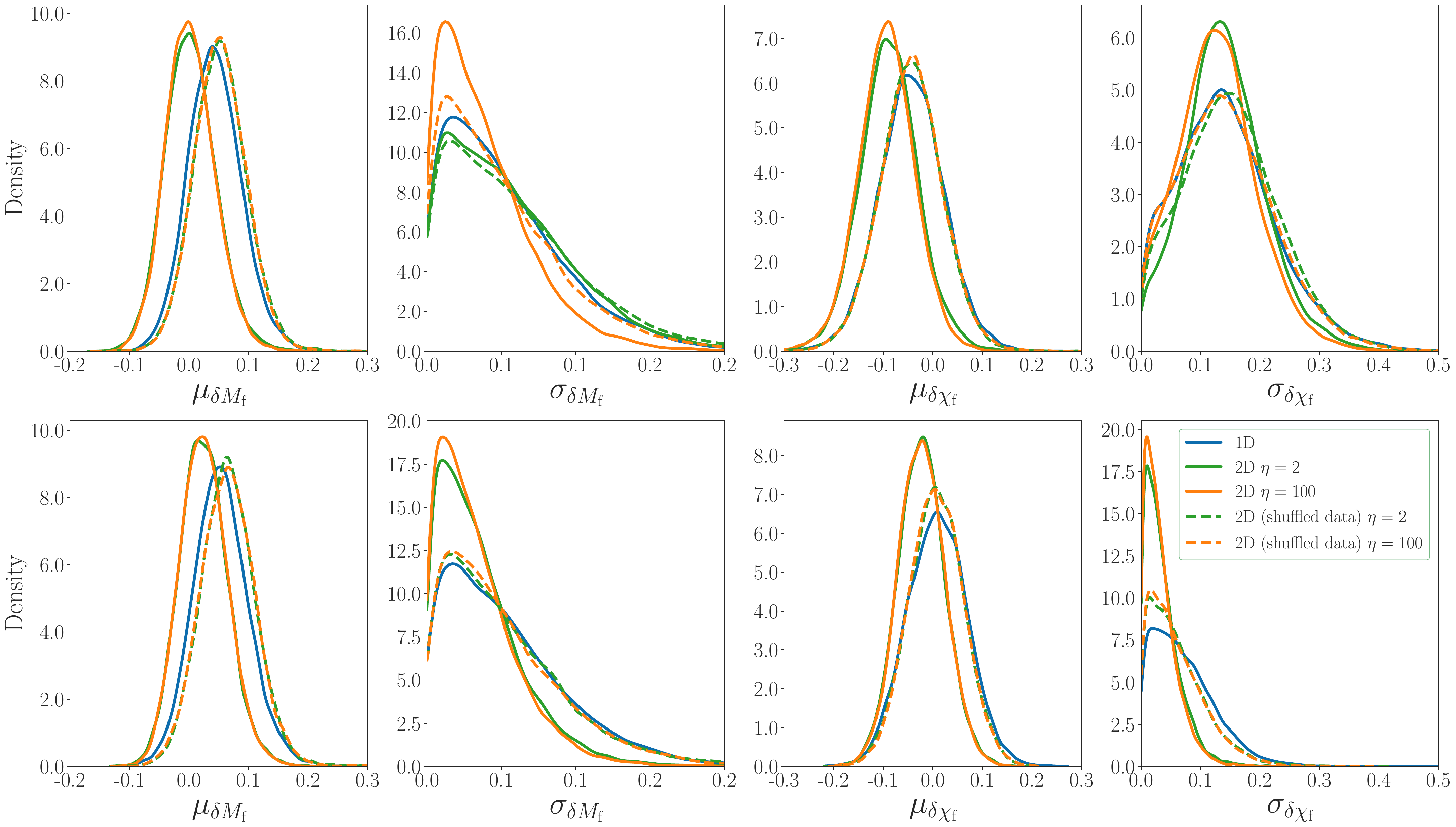}
    \caption{Hyperparameter posterior distributions derived from GWTC IMR analyses with varying configurations with (left) and without (right) GW190814. The blue curves correspond to 1D analyses, while the orange and green curves show results from 2D analyses, with priors of  $\eta=2$ and $\eta=100$ respectively. Dashed curves show 2D results from shuffled individual-event likelihoods (App.~\ref{sec:sanity checks}). When the individual-event likelihoods are informative, hyperposteriors are not sensitive to the choice of $\eta$ (solid orange versus green orange); 1D results can be recovered by erasing correlation information from the individual-event likelihoods through sample shuffling (dashed orange and green versus solid blue).}
    \label{fig:sanity check}
\end{figure*}

We repeat the hierarchical analysis and show results in Fig.~\ref{fig:sanity check}. Each subplot shows hyperparamater posterior distributions from analyses with varying configurations. The blue curves show the results from 1D analyses, while other curves give results from 2D analyses. Orange and green curves correspond to the prior $\eta=2$ and $\eta=100$ cases, respectively. Dashed curves are results of 2D analyses on shuffled samples. 
Comparing the orange and green curves indicates that varying the prior $\eta$ has minimal impact on the resulting posterior distributions. 
When the samples are shuffled, the 1D results and 2D results are identical. 

\section{Full 4D results}\label{appendixC}

\begin{figure*}[]
    \centering
\includegraphics[width=1.0\textwidth]{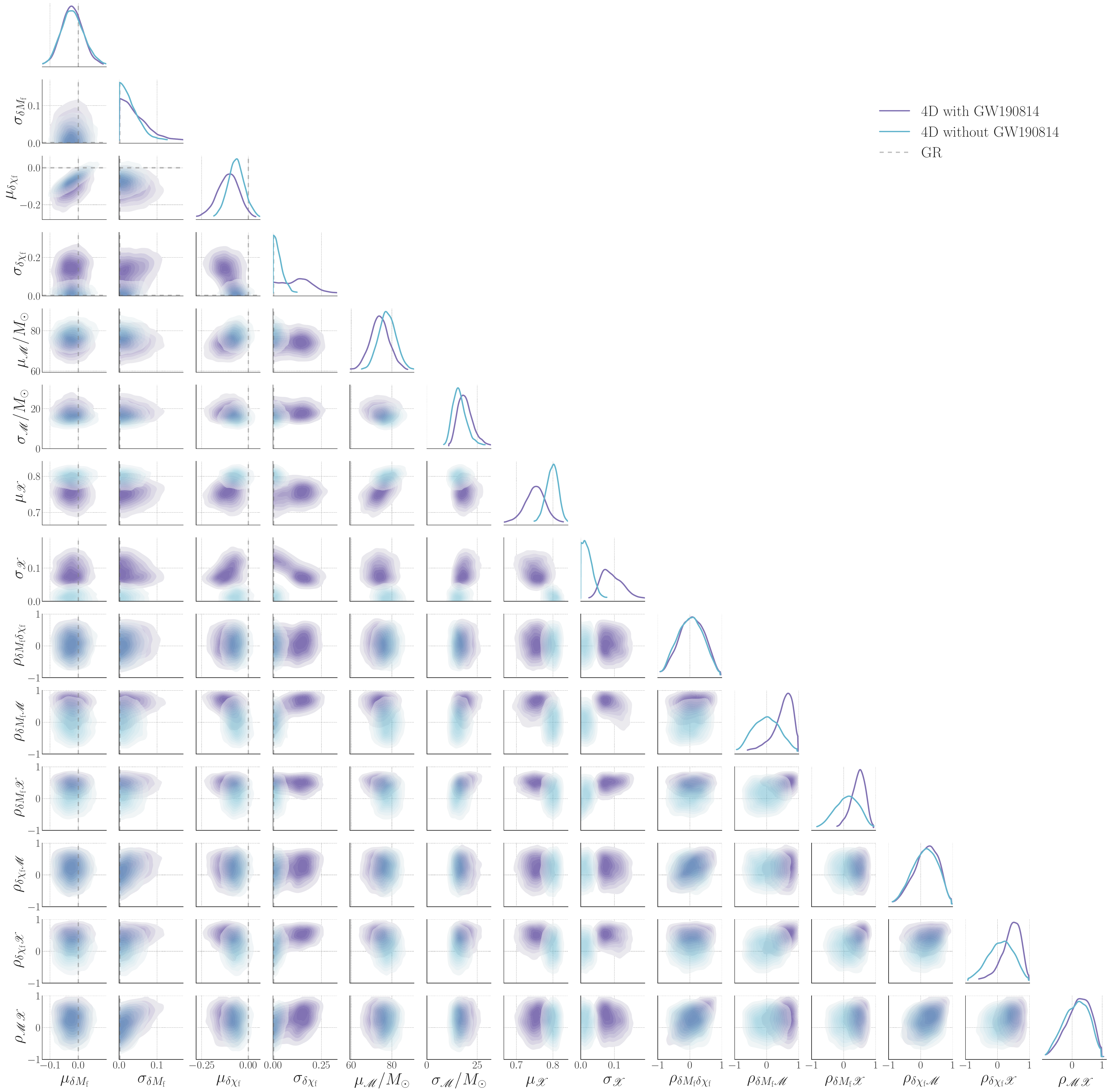}
    \caption{Posterior distributions of all hyperparameters of the 4D IMR analysis of GWTC-3, including (purple) and excluding (light blue) GW190814.
    Contours enclose increments of 10\% of probability mass starting at 90\% for the outermost contour. The purple (light blue) distribution here is the same as the green distribution in Fig.~\ref{fig:4D} (Fig.~\ref{fig:4D_woGW190814}). See Sec.~\ref{sec:realdata:4d} in the main text for a discussion of these results.
    }
    \label{fig:14corner}
\end{figure*}

In this appendix we show corner plots for all hyperparameters of the full 4D analysis with (purple) and without (light blue) GW190814 in Fig.~\ref{fig:14corner}.

\bibliography{reference}

\begin{thebibliography}{67}%
\makeatletter
\providecommand \@ifxundefined [1]{%
 \@ifx{#1\undefined}
}%
\providecommand \@ifnum [1]{%
 \ifnum #1\expandafter \@firstoftwo
 \else \expandafter \@secondoftwo
 \fi
}%
\providecommand \@ifx [1]{%
 \ifx #1\expandafter \@firstoftwo
 \else \expandafter \@secondoftwo
 \fi
}%
\providecommand \natexlab [1]{#1}%
\providecommand \enquote  [1]{``#1''}%
\providecommand \bibnamefont  [1]{#1}%
\providecommand \bibfnamefont [1]{#1}%
\providecommand \citenamefont [1]{#1}%
\providecommand \href@noop [0]{\@secondoftwo}%
\providecommand \href [0]{\begingroup \@sanitize@url \@href}%
\providecommand \@href[1]{\@@startlink{#1}\@@href}%
\providecommand \@@href[1]{\endgroup#1\@@endlink}%
\providecommand \@sanitize@url [0]{\catcode `\\12\catcode `\$12\catcode
  `\&12\catcode `\#12\catcode `\^12\catcode `\_12\catcode `\%12\relax}%
\providecommand \@@startlink[1]{}%
\providecommand \@@endlink[0]{}%
\providecommand \url  [0]{\begingroup\@sanitize@url \@url }%
\providecommand \@url [1]{\endgroup\@href {#1}{\urlprefix }}%
\providecommand \urlprefix  [0]{URL }%
\providecommand \Eprint [0]{\href }%
\providecommand \doibase [0]{https://doi.org/}%
\providecommand \selectlanguage [0]{\@gobble}%
\providecommand \bibinfo  [0]{\@secondoftwo}%
\providecommand \bibfield  [0]{\@secondoftwo}%
\providecommand \translation [1]{[#1]}%
\providecommand \BibitemOpen [0]{}%
\providecommand \bibitemStop [0]{}%
\providecommand \bibitemNoStop [0]{.\EOS\space}%
\providecommand \EOS [0]{\spacefactor3000\relax}%
\providecommand \BibitemShut  [1]{\csname bibitem#1\endcsname}%
\let\auto@bib@innerbib\@empty
\bibitem [{\citenamefont {Aasi}\ \emph {et~al.}(2015)\citenamefont {Aasi} \emph
  {et~al.}}]{AdLIGO}%
  \BibitemOpen
  \bibfield  {author} {\bibinfo {author} {\bibfnamefont {J.}~\bibnamefont
  {Aasi}} \emph {et~al.} (\bibinfo {collaboration} {LIGO Scientific}),\
  }\bibfield  {title} {\bibinfo {title} {{Advanced LIGO}},\ }\href
  {https://doi.org/10.1088/0264-9381/32/7/074001} {\bibfield  {journal}
  {\bibinfo  {journal} {Class. Quant. Grav.}\ }\textbf {\bibinfo {volume}
  {32}},\ \bibinfo {pages} {074001} (\bibinfo {year} {2015})},\ \Eprint
  {https://arxiv.org/abs/1411.4547} {arXiv:1411.4547 [gr-qc]} \BibitemShut
  {NoStop}%
\bibitem [{\citenamefont {Acernese}\ \emph {et~al.}(2015)\citenamefont
  {Acernese} \emph {et~al.}}]{AdVirgo}%
  \BibitemOpen
  \bibfield  {author} {\bibinfo {author} {\bibfnamefont {F.}~\bibnamefont
  {Acernese}} \emph {et~al.} (\bibinfo {collaboration} {VIRGO}),\ }\bibfield
  {title} {\bibinfo {title} {{Advanced Virgo: a second-generation
  interferometric gravitational wave detector}},\ }\href
  {https://doi.org/10.1088/0264-9381/32/2/024001} {\bibfield  {journal}
  {\bibinfo  {journal} {Class. Quant. Grav.}\ }\textbf {\bibinfo {volume}
  {32}},\ \bibinfo {pages} {024001} (\bibinfo {year} {2015})},\ \Eprint
  {https://arxiv.org/abs/1408.3978} {arXiv:1408.3978 [gr-qc]} \BibitemShut
  {NoStop}%
\bibitem [{\citenamefont {Abbott}\ \emph
  {et~al.}(2019{\natexlab{a}})\citenamefont {Abbott} \emph {et~al.}}]{GWTC1}%
  \BibitemOpen
  \bibfield  {author} {\bibinfo {author} {\bibfnamefont {B.~P.}\ \bibnamefont
  {Abbott}} \emph {et~al.} (\bibinfo {collaboration} {LIGO Scientific,
  Virgo}),\ }\bibfield  {title} {\bibinfo {title} {{GWTC-1: A
  Gravitational-Wave Transient Catalog of Compact Binary Mergers Observed by
  LIGO and Virgo during the First and Second Observing Runs}},\ }\href
  {https://doi.org/10.1103/PhysRevX.9.031040} {\bibfield  {journal} {\bibinfo
  {journal} {Phys. Rev. X}\ }\textbf {\bibinfo {volume} {9}},\ \bibinfo {pages}
  {031040} (\bibinfo {year} {2019}{\natexlab{a}})},\ \Eprint
  {https://arxiv.org/abs/1811.12907} {arXiv:1811.12907 [astro-ph.HE]}
  \BibitemShut {NoStop}%
\bibitem [{\citenamefont {Abbott}\ \emph
  {et~al.}(2021{\natexlab{a}})\citenamefont {Abbott} \emph {et~al.}}]{GWTC2}%
  \BibitemOpen
  \bibfield  {author} {\bibinfo {author} {\bibfnamefont {R.}~\bibnamefont
  {Abbott}} \emph {et~al.} (\bibinfo {collaboration} {LIGO Scientific,
  Virgo}),\ }\bibfield  {title} {\bibinfo {title} {{GWTC-2: Compact Binary
  Coalescences Observed by LIGO and Virgo During the First Half of the Third
  Observing Run}},\ }\href {https://doi.org/10.1103/PhysRevX.11.021053}
  {\bibfield  {journal} {\bibinfo  {journal} {Phys. Rev. X}\ }\textbf {\bibinfo
  {volume} {11}},\ \bibinfo {pages} {021053} (\bibinfo {year}
  {2021}{\natexlab{a}})},\ \Eprint {https://arxiv.org/abs/2010.14527}
  {arXiv:2010.14527 [gr-qc]} \BibitemShut {NoStop}%
\bibitem [{\citenamefont {Abbott}\ \emph {et~al.}({\natexlab{a}})\citenamefont
  {Abbott} \emph {et~al.}}]{GWTC2.1}%
  \BibitemOpen
  \bibfield  {author} {\bibinfo {author} {\bibfnamefont {R.}~\bibnamefont
  {Abbott}} \emph {et~al.} (\bibinfo {collaboration} {LIGO Scientific,
  VIRGO}),\ }\bibfield  {title} {\bibinfo {title} {{GWTC-2.1: Deep Extended
  Catalog of Compact Binary Coalescences Observed by LIGO and Virgo During the
  First Half of the Third Observing Run}},\ }\href@noop {} {\
  ({\natexlab{a}})},\ \Eprint {https://arxiv.org/abs/2108.01045}
  {arXiv:2108.01045 [gr-qc]} \BibitemShut {NoStop}%
\bibitem [{\citenamefont {Abbott}\ \emph {et~al.}({\natexlab{b}})\citenamefont
  {Abbott} \emph {et~al.}}]{GWTC3}%
  \BibitemOpen
  \bibfield  {author} {\bibinfo {author} {\bibfnamefont {R.}~\bibnamefont
  {Abbott}} \emph {et~al.} (\bibinfo {collaboration} {LIGO Scientific, VIRGO,
  KAGRA}),\ }\bibfield  {title} {\bibinfo {title} {{GWTC-3: Compact Binary
  Coalescences Observed by LIGO and Virgo During the Second Part of the Third
  Observing Run}},\ }\href@noop {} {\  ({\natexlab{b}})},\ \Eprint
  {https://arxiv.org/abs/2111.03606} {arXiv:2111.03606 [gr-qc]} \BibitemShut
  {NoStop}%
\bibitem [{\citenamefont {Abbott}\ \emph
  {et~al.}(2019{\natexlab{b}})\citenamefont {Abbott} \emph
  {et~al.}}]{GWTC1_BBH}%
  \BibitemOpen
  \bibfield  {author} {\bibinfo {author} {\bibfnamefont {B.~P.}\ \bibnamefont
  {Abbott}} \emph {et~al.} (\bibinfo {collaboration} {LIGO Scientific,
  Virgo}),\ }\bibfield  {title} {\bibinfo {title} {{Binary Black Hole
  Population Properties Inferred from the First and Second Observing Runs of
  Advanced LIGO and Advanced Virgo}},\ }\href
  {https://doi.org/10.3847/2041-8213/ab3800} {\bibfield  {journal} {\bibinfo
  {journal} {Astrophys. J. Lett.}\ }\textbf {\bibinfo {volume} {882}},\
  \bibinfo {pages} {L24} (\bibinfo {year} {2019}{\natexlab{b}})},\ \Eprint
  {https://arxiv.org/abs/1811.12940} {arXiv:1811.12940 [astro-ph.HE]}
  \BibitemShut {NoStop}%
\bibitem [{\citenamefont {Abbott}\ \emph
  {et~al.}(2021{\natexlab{b}})\citenamefont {Abbott} \emph
  {et~al.}}]{GWTC1_Hubble}%
  \BibitemOpen
  \bibfield  {author} {\bibinfo {author} {\bibfnamefont {B.~P.}\ \bibnamefont
  {Abbott}} \emph {et~al.} (\bibinfo {collaboration} {LIGO Scientific, Virgo,
  VIRGO}),\ }\bibfield  {title} {\bibinfo {title} {{A Gravitational-wave
  Measurement of the Hubble Constant Following the Second Observing Run of
  Advanced LIGO and Virgo}},\ }\href {https://doi.org/10.3847/1538-4357/abdcb7}
  {\bibfield  {journal} {\bibinfo  {journal} {Astrophys. J.}\ }\textbf
  {\bibinfo {volume} {909}},\ \bibinfo {pages} {218} (\bibinfo {year}
  {2021}{\natexlab{b}})},\ \Eprint {https://arxiv.org/abs/1908.06060}
  {arXiv:1908.06060 [astro-ph.CO]} \BibitemShut {NoStop}%
\bibitem [{\citenamefont {Abbott}\ \emph
  {et~al.}(2021{\natexlab{c}})\citenamefont {Abbott} \emph
  {et~al.}}]{GWTC2_CBC}%
  \BibitemOpen
  \bibfield  {author} {\bibinfo {author} {\bibfnamefont {R.}~\bibnamefont
  {Abbott}} \emph {et~al.} (\bibinfo {collaboration} {LIGO Scientific,
  Virgo}),\ }\bibfield  {title} {\bibinfo {title} {{Population Properties of
  Compact Objects from the Second LIGO-Virgo Gravitational-Wave Transient
  Catalog}},\ }\href {https://doi.org/10.3847/2041-8213/abe949} {\bibfield
  {journal} {\bibinfo  {journal} {Astrophys. J. Lett.}\ }\textbf {\bibinfo
  {volume} {913}},\ \bibinfo {pages} {L7} (\bibinfo {year}
  {2021}{\natexlab{c}})},\ \Eprint {https://arxiv.org/abs/2010.14533}
  {arXiv:2010.14533 [astro-ph.HE]} \BibitemShut {NoStop}%
\bibitem [{\citenamefont {Abbott}\ \emph
  {et~al.}(2021{\natexlab{d}})\citenamefont {Abbott} \emph
  {et~al.}}]{GWTC2_Gamma}%
  \BibitemOpen
  \bibfield  {author} {\bibinfo {author} {\bibfnamefont {R.}~\bibnamefont
  {Abbott}} \emph {et~al.} (\bibinfo {collaboration} {LIGO Scientific,
  Virgo}),\ }\bibfield  {title} {\bibinfo {title} {{Search for Gravitational
  Waves Associated with Gamma-Ray Bursts Detected by Fermi and Swift During the
  LIGO-Virgo Run O3a}},\ }\href {https://doi.org/10.3847/1538-4357/abee15}
  {\bibfield  {journal} {\bibinfo  {journal} {Astrophys. J.}\ }\textbf
  {\bibinfo {volume} {915}},\ \bibinfo {pages} {86} (\bibinfo {year}
  {2021}{\natexlab{d}})},\ \Eprint {https://arxiv.org/abs/2010.14550}
  {arXiv:2010.14550 [astro-ph.HE]} \BibitemShut {NoStop}%
\bibitem [{\citenamefont {Abbott}\ \emph
  {et~al.}(2021{\natexlab{e}})\citenamefont {Abbott} \emph
  {et~al.}}]{GWTC2_Lensing}%
  \BibitemOpen
  \bibfield  {author} {\bibinfo {author} {\bibfnamefont {R.}~\bibnamefont
  {Abbott}} \emph {et~al.} (\bibinfo {collaboration} {LIGO Scientific,
  VIRGO}),\ }\bibfield  {title} {\bibinfo {title} {{Search for Lensing
  Signatures in the Gravitational-Wave Observations from the First Half of
  LIGO\textendash{}Virgo\textquoteright{}s Third Observing Run}},\ }\href
  {https://doi.org/10.3847/1538-4357/ac23db} {\bibfield  {journal} {\bibinfo
  {journal} {Astrophys. J.}\ }\textbf {\bibinfo {volume} {923}},\ \bibinfo
  {pages} {14} (\bibinfo {year} {2021}{\natexlab{e}})},\ \Eprint
  {https://arxiv.org/abs/2105.06384} {arXiv:2105.06384 [gr-qc]} \BibitemShut
  {NoStop}%
\bibitem [{\citenamefont {Abbott}\ \emph
  {et~al.}(2023{\natexlab{a}})\citenamefont {Abbott} \emph
  {et~al.}}]{GWTC3_CBC}%
  \BibitemOpen
  \bibfield  {author} {\bibinfo {author} {\bibfnamefont {R.}~\bibnamefont
  {Abbott}} \emph {et~al.} (\bibinfo {collaboration} {KAGRA, VIRGO, LIGO
  Scientific}),\ }\bibfield  {title} {\bibinfo {title} {{Population of Merging
  Compact Binaries Inferred Using Gravitational Waves through GWTC-3}},\ }\href
  {https://doi.org/10.1103/PhysRevX.13.011048} {\bibfield  {journal} {\bibinfo
  {journal} {Phys. Rev. X}\ }\textbf {\bibinfo {volume} {13}},\ \bibinfo
  {pages} {011048} (\bibinfo {year} {2023}{\natexlab{a}})},\ \Eprint
  {https://arxiv.org/abs/2111.03634} {arXiv:2111.03634 [astro-ph.HE]}
  \BibitemShut {NoStop}%
\bibitem [{\citenamefont {Abbott}\ \emph
  {et~al.}(2023{\natexlab{b}})\citenamefont {Abbott} \emph
  {et~al.}}]{GWTC3_Cosmo}%
  \BibitemOpen
  \bibfield  {author} {\bibinfo {author} {\bibfnamefont {R.}~\bibnamefont
  {Abbott}} \emph {et~al.} (\bibinfo {collaboration} {LIGO Scientific, VIRGO,
  KAGRA}),\ }\bibfield  {title} {\bibinfo {title} {{Constraints on the Cosmic
  Expansion History from GWTC\textendash{}3}},\ }\href
  {https://doi.org/10.3847/1538-4357/ac74bb} {\bibfield  {journal} {\bibinfo
  {journal} {Astrophys. J.}\ }\textbf {\bibinfo {volume} {949}},\ \bibinfo
  {pages} {76} (\bibinfo {year} {2023}{\natexlab{b}})},\ \Eprint
  {https://arxiv.org/abs/2111.03604} {arXiv:2111.03604 [astro-ph.CO]}
  \BibitemShut {NoStop}%
\bibitem [{\citenamefont {Abbott}\ \emph {et~al.}(2022)\citenamefont {Abbott}
  \emph {et~al.}}]{GWTC3_Gamma}%
  \BibitemOpen
  \bibfield  {author} {\bibinfo {author} {\bibfnamefont {R.}~\bibnamefont
  {Abbott}} \emph {et~al.} (\bibinfo {collaboration} {LIGO Scientific, VIRGO,
  KAGRA}),\ }\bibfield  {title} {\bibinfo {title} {{Search for Gravitational
  Waves Associated with Gamma-Ray Bursts Detected by Fermi and Swift during the
  LIGO\textendash{}Virgo Run O3b}},\ }\href
  {https://doi.org/10.3847/1538-4357/ac532b} {\bibfield  {journal} {\bibinfo
  {journal} {Astrophys. J.}\ }\textbf {\bibinfo {volume} {928}},\ \bibinfo
  {pages} {186} (\bibinfo {year} {2022})},\ \Eprint
  {https://arxiv.org/abs/2111.03608} {arXiv:2111.03608 [astro-ph.HE]}
  \BibitemShut {NoStop}%
\bibitem [{\citenamefont {Abbott}\ \emph
  {et~al.}(2019{\natexlab{c}})\citenamefont {Abbott} \emph
  {et~al.}}]{GWTC1-TGR}%
  \BibitemOpen
  \bibfield  {author} {\bibinfo {author} {\bibfnamefont {B.~P.}\ \bibnamefont
  {Abbott}} \emph {et~al.} (\bibinfo {collaboration} {LIGO Scientific,
  Virgo}),\ }\bibfield  {title} {\bibinfo {title} {{Tests of General Relativity
  with the Binary Black Hole Signals from the LIGO-Virgo Catalog GWTC-1}},\
  }\href {https://doi.org/10.1103/PhysRevD.100.104036} {\bibfield  {journal}
  {\bibinfo  {journal} {Phys. Rev. D}\ }\textbf {\bibinfo {volume} {100}},\
  \bibinfo {pages} {104036} (\bibinfo {year} {2019}{\natexlab{c}})},\ \Eprint
  {https://arxiv.org/abs/1903.04467} {arXiv:1903.04467 [gr-qc]} \BibitemShut
  {NoStop}%
\bibitem [{\citenamefont {Abbott}\ \emph
  {et~al.}(2021{\natexlab{f}})\citenamefont {Abbott} \emph
  {et~al.}}]{GWTC2-TGR}%
  \BibitemOpen
  \bibfield  {author} {\bibinfo {author} {\bibfnamefont {R.}~\bibnamefont
  {Abbott}} \emph {et~al.} (\bibinfo {collaboration} {LIGO Scientific,
  Virgo}),\ }\bibfield  {title} {\bibinfo {title} {{Tests of general relativity
  with binary black holes from the second LIGO-Virgo gravitational-wave
  transient catalog}},\ }\href {https://doi.org/10.1103/PhysRevD.103.122002}
  {\bibfield  {journal} {\bibinfo  {journal} {Phys. Rev. D}\ }\textbf {\bibinfo
  {volume} {103}},\ \bibinfo {pages} {122002} (\bibinfo {year}
  {2021}{\natexlab{f}})},\ \Eprint {https://arxiv.org/abs/2010.14529}
  {arXiv:2010.14529 [gr-qc]} \BibitemShut {NoStop}%
\bibitem [{\citenamefont {Abbott}\ \emph
  {et~al.}(2021{\natexlab{g}})\citenamefont {Abbott} \emph
  {et~al.}}]{GWTC3-TGR}%
  \BibitemOpen
  \bibfield  {author} {\bibinfo {author} {\bibfnamefont {R.}~\bibnamefont
  {Abbott}} \emph {et~al.} (\bibinfo {collaboration} {LIGO Scientific, VIRGO,
  KAGRA}),\ }\bibfield  {title} {\bibinfo {title} {{Tests of General Relativity
  with GWTC-3}},\ }\href@noop {} {\  (\bibinfo {year} {2021}{\natexlab{g}})},\
  \Eprint {https://arxiv.org/abs/2112.06861} {arXiv:2112.06861 [gr-qc]}
  \BibitemShut {NoStop}%
\bibitem [{\citenamefont {Zimmerman}\ \emph {et~al.}(2019)\citenamefont
  {Zimmerman}, \citenamefont {Haster},\ and\ \citenamefont
  {Chatziioannou}}]{Zimmerman:2019wzo}%
  \BibitemOpen
  \bibfield  {author} {\bibinfo {author} {\bibfnamefont {A.}~\bibnamefont
  {Zimmerman}}, \bibinfo {author} {\bibfnamefont {C.-J.}\ \bibnamefont
  {Haster}},\ and\ \bibinfo {author} {\bibfnamefont {K.}~\bibnamefont
  {Chatziioannou}},\ }\bibfield  {title} {\bibinfo {title} {{On combining
  information from multiple gravitational wave sources}},\ }\href
  {https://doi.org/10.1103/PhysRevD.99.124044} {\bibfield  {journal} {\bibinfo
  {journal} {Phys. Rev. D}\ }\textbf {\bibinfo {volume} {99}},\ \bibinfo
  {pages} {124044} (\bibinfo {year} {2019})},\ \Eprint
  {https://arxiv.org/abs/1903.11008} {arXiv:1903.11008 [astro-ph.IM]}
  \BibitemShut {NoStop}%
\bibitem [{\citenamefont {Isi}\ \emph {et~al.}(2019{\natexlab{a}})\citenamefont
  {Isi}, \citenamefont {Chatziioannou},\ and\ \citenamefont {Farr}}]{Hier}%
  \BibitemOpen
  \bibfield  {author} {\bibinfo {author} {\bibfnamefont {M.}~\bibnamefont
  {Isi}}, \bibinfo {author} {\bibfnamefont {K.}~\bibnamefont {Chatziioannou}},\
  and\ \bibinfo {author} {\bibfnamefont {W.~M.}\ \bibnamefont {Farr}},\
  }\bibfield  {title} {\bibinfo {title} {{Hierarchical test of general
  relativity with gravitational waves}},\ }\href
  {https://doi.org/10.1103/PhysRevLett.123.121101} {\bibfield  {journal}
  {\bibinfo  {journal} {Phys. Rev. Lett.}\ }\textbf {\bibinfo {volume} {123}},\
  \bibinfo {pages} {121101} (\bibinfo {year} {2019}{\natexlab{a}})},\ \Eprint
  {https://arxiv.org/abs/1904.08011} {arXiv:1904.08011 [gr-qc]} \BibitemShut
  {NoStop}%
\bibitem [{\citenamefont {Isi}\ \emph {et~al.}(2022)\citenamefont {Isi},
  \citenamefont {Farr},\ and\ \citenamefont {Chatziioannou}}]{Isi:2022cii}%
  \BibitemOpen
  \bibfield  {author} {\bibinfo {author} {\bibfnamefont {M.}~\bibnamefont
  {Isi}}, \bibinfo {author} {\bibfnamefont {W.~M.}\ \bibnamefont {Farr}},\ and\
  \bibinfo {author} {\bibfnamefont {K.}~\bibnamefont {Chatziioannou}},\
  }\bibfield  {title} {\bibinfo {title} {{Comparing Bayes factors and
  hierarchical inference for testing general relativity with gravitational
  waves}},\ }\href {https://doi.org/10.1103/PhysRevD.106.024048} {\bibfield
  {journal} {\bibinfo  {journal} {Phys. Rev. D}\ }\textbf {\bibinfo {volume}
  {106}},\ \bibinfo {pages} {024048} (\bibinfo {year} {2022})},\ \Eprint
  {https://arxiv.org/abs/2204.10742} {arXiv:2204.10742 [gr-qc]} \BibitemShut
  {NoStop}%
\bibitem [{\citenamefont {Pacilio}\ \emph {et~al.}(2024)\citenamefont
  {Pacilio}, \citenamefont {Gerosa},\ and\ \citenamefont
  {Bhagwat}}]{Pacilio:2023uef}%
  \BibitemOpen
  \bibfield  {author} {\bibinfo {author} {\bibfnamefont {C.}~\bibnamefont
  {Pacilio}}, \bibinfo {author} {\bibfnamefont {D.}~\bibnamefont {Gerosa}},\
  and\ \bibinfo {author} {\bibfnamefont {S.}~\bibnamefont {Bhagwat}},\
  }\bibfield  {title} {\bibinfo {title} {{Catalog variance of testing general
  relativity with gravitational-wave data}},\ }\href
  {https://doi.org/10.1103/PhysRevD.109.L081302} {\bibfield  {journal}
  {\bibinfo  {journal} {Phys. Rev. D}\ }\textbf {\bibinfo {volume} {109}},\
  \bibinfo {pages} {L081302} (\bibinfo {year} {2024})},\ \Eprint
  {https://arxiv.org/abs/2310.03811} {arXiv:2310.03811 [gr-qc]} \BibitemShut
  {NoStop}%
\bibitem [{\citenamefont {Payne}\ \emph {et~al.}(2023)\citenamefont {Payne},
  \citenamefont {Isi}, \citenamefont {Chatziioannou},\ and\ \citenamefont
  {Farr}}]{Payne:2023kwj}%
  \BibitemOpen
  \bibfield  {author} {\bibinfo {author} {\bibfnamefont {E.}~\bibnamefont
  {Payne}}, \bibinfo {author} {\bibfnamefont {M.}~\bibnamefont {Isi}}, \bibinfo
  {author} {\bibfnamefont {K.}~\bibnamefont {Chatziioannou}},\ and\ \bibinfo
  {author} {\bibfnamefont {W.~M.}\ \bibnamefont {Farr}},\ }\bibfield  {title}
  {\bibinfo {title} {{Fortifying gravitational-wave tests of general relativity
  against astrophysical assumptions}},\ }\Eprint
  {https://arxiv.org/abs/2309.04528} {arXiv:2309.04528 [gr-qc]}  (\bibinfo
  {year} {2023})\BibitemShut {NoStop}%
\bibitem [{\citenamefont {Magee}\ \emph {et~al.}(2024)\citenamefont {Magee},
  \citenamefont {Isi}, \citenamefont {Payne}, \citenamefont {Chatziioannou},
  \citenamefont {Farr}, \citenamefont {Pratten},\ and\ \citenamefont
  {Vitale}}]{Magee:2023muf}%
  \BibitemOpen
  \bibfield  {author} {\bibinfo {author} {\bibfnamefont {R.}~\bibnamefont
  {Magee}}, \bibinfo {author} {\bibfnamefont {M.}~\bibnamefont {Isi}}, \bibinfo
  {author} {\bibfnamefont {E.}~\bibnamefont {Payne}}, \bibinfo {author}
  {\bibfnamefont {K.}~\bibnamefont {Chatziioannou}}, \bibinfo {author}
  {\bibfnamefont {W.~M.}\ \bibnamefont {Farr}}, \bibinfo {author}
  {\bibfnamefont {G.}~\bibnamefont {Pratten}},\ and\ \bibinfo {author}
  {\bibfnamefont {S.}~\bibnamefont {Vitale}},\ }\bibfield  {title} {\bibinfo
  {title} {{Impact of selection biases on tests of general relativity with
  gravitational-wave inspirals}},\ }\href
  {https://doi.org/10.1103/PhysRevD.109.023014} {\bibfield  {journal} {\bibinfo
   {journal} {Phys. Rev. D}\ }\textbf {\bibinfo {volume} {109}},\ \bibinfo
  {pages} {023014} (\bibinfo {year} {2024})},\ \Eprint
  {https://arxiv.org/abs/2311.03656} {arXiv:2311.03656 [gr-qc]} \BibitemShut
  {NoStop}%
\bibitem [{\citenamefont {Essick}\ and\ \citenamefont
  {Fishbach}(2023)}]{Essick:2023upv}%
  \BibitemOpen
  \bibfield  {author} {\bibinfo {author} {\bibfnamefont {R.}~\bibnamefont
  {Essick}}\ and\ \bibinfo {author} {\bibfnamefont {M.}~\bibnamefont
  {Fishbach}},\ }\bibfield  {title} {\bibinfo {title} {{DAGnabbit! Ensuring
  Consistency between Noise and Detection in Hierarchical Bayesian
  Inference}},\ }\Eprint {https://arxiv.org/abs/2310.02017} {arXiv:2310.02017
  [gr-qc]}  (\bibinfo {year} {2023})\BibitemShut {NoStop}%
\bibitem [{\citenamefont {James}\ and\ \citenamefont
  {Stein}(1961)}]{James:1961}%
  \BibitemOpen
  \bibfield  {author} {\bibinfo {author} {\bibfnamefont {W.}~\bibnamefont
  {James}}\ and\ \bibinfo {author} {\bibfnamefont {C.}~\bibnamefont {Stein}},\
  }\bibfield  {title} {\bibinfo {title} {Estimation with quadratic loss},\ }in\
  \href@noop {} {\emph {\bibinfo {booktitle} {Proc. 4th {B}erkeley {S}ympos.
  {M}ath. {S}tatist. and {P}rob., {V}ol. {I}}}}\ (\bibinfo  {publisher} {Univ.
  California Press, Berkeley, Calif.},\ \bibinfo {year} {1961})\ pp.\ \bibinfo
  {pages} {361--379}\BibitemShut {NoStop}%
\bibitem [{\citenamefont {Lindley}\ and\ \citenamefont
  {Smith}(1972)}]{Lindley:1972}%
  \BibitemOpen
  \bibfield  {author} {\bibinfo {author} {\bibfnamefont {D.~V.}\ \bibnamefont
  {Lindley}}\ and\ \bibinfo {author} {\bibfnamefont {A.~F.~M.}\ \bibnamefont
  {Smith}},\ }\bibfield  {title} {\bibinfo {title} {Bayes estimates for the
  linear model},\ }\href {http://www.jstor.org/stable/2985048} {\bibfield
  {journal} {\bibinfo  {journal} {Journal of the Royal Statistical Society.
  Series B (Methodological)}\ }\textbf {\bibinfo {volume} {34}},\ \bibinfo
  {pages} {1} (\bibinfo {year} {1972})}\BibitemShut {NoStop}%
\bibitem [{\citenamefont {{Efron}}\ and\ \citenamefont
  {{Morris}}(1977)}]{Efron:1977}%
  \BibitemOpen
  \bibfield  {author} {\bibinfo {author} {\bibfnamefont {B.}~\bibnamefont
  {{Efron}}}\ and\ \bibinfo {author} {\bibfnamefont {C.}~\bibnamefont
  {{Morris}}},\ }\bibfield  {title} {\bibinfo {title} {{Stein's Paradox in
  Statistics}},\ }\href {https://doi.org/10.1038/scientificamerican0577-119}
  {\bibfield  {journal} {\bibinfo  {journal} {Scientific American}\ }\textbf
  {\bibinfo {volume} {236}},\ \bibinfo {pages} {119} (\bibinfo {year}
  {1977})}\BibitemShut {NoStop}%
\bibitem [{\citenamefont {Rubin}(1981)}]{Rubin:1981}%
  \BibitemOpen
  \bibfield  {author} {\bibinfo {author} {\bibfnamefont {D.~B.}\ \bibnamefont
  {Rubin}},\ }\bibfield  {title} {\bibinfo {title} {Estimation in parallel
  randomized experiments},\ }\href {http://www.jstor.org/stable/1164617}
  {\bibfield  {journal} {\bibinfo  {journal} {Journal of Educational
  Statistics}\ }\textbf {\bibinfo {volume} {6}},\ \bibinfo {pages} {377}
  (\bibinfo {year} {1981})}\BibitemShut {NoStop}%
\bibitem [{\citenamefont {Thrane}\ and\ \citenamefont
  {Talbot}(2019)}]{Thrane:2018qnx}%
  \BibitemOpen
  \bibfield  {author} {\bibinfo {author} {\bibfnamefont {E.}~\bibnamefont
  {Thrane}}\ and\ \bibinfo {author} {\bibfnamefont {C.}~\bibnamefont
  {Talbot}},\ }\bibfield  {title} {\bibinfo {title} {{An introduction to
  Bayesian inference in gravitational-wave astronomy: parameter estimation,
  model selection, and hierarchical models}},\ }\href
  {https://doi.org/10.1017/pasa.2019.2} {\bibfield  {journal} {\bibinfo
  {journal} {Publ. Astron. Soc. Austral.}\ }\textbf {\bibinfo {volume} {36}},\
  \bibinfo {pages} {e010} (\bibinfo {year} {2019})},\ \bibinfo {note}
  {[Erratum: Publ.Astron.Soc.Austral. 37, e036 (2020)]},\ \Eprint
  {https://arxiv.org/abs/1809.02293} {arXiv:1809.02293 [astro-ph.IM]}
  \BibitemShut {NoStop}%
\bibitem [{\citenamefont {Vitale}\ \emph {et~al.}(2022)\citenamefont {Vitale},
  \citenamefont {Gerosa}, \citenamefont {Farr},\ and\ \citenamefont
  {Taylor}}]{Vitale:2020aaz}%
  \BibitemOpen
  \bibfield  {author} {\bibinfo {author} {\bibfnamefont {S.}~\bibnamefont
  {Vitale}}, \bibinfo {author} {\bibfnamefont {D.}~\bibnamefont {Gerosa}},
  \bibinfo {author} {\bibfnamefont {W.}~\bibnamefont {Farr}},\ and\ \bibinfo
  {author} {\bibfnamefont {S.}~\bibnamefont {Taylor}},\ }\bibinfo {title}
  {Inferring the properties of a population of compact binaries in presence of
  selection effects},\ in\ \href
  {https://doi.org/10.1007/978-981-15-4702-7_45-1} {\emph {\bibinfo {booktitle}
  {Handbook of Gravitational Wave Astronomy}}}\ (\bibinfo  {publisher}
  {Springer, Singapore},\ \bibinfo {year} {2022})\ \Eprint
  {https://arxiv.org/abs/2007.05579} {arXiv:2007.05579 [astro-ph.IM]}
  \BibitemShut {NoStop}%
\bibitem [{\citenamefont {Yunes}\ and\ \citenamefont
  {Pretorius}(2009)}]{Yunes:2009ke}%
  \BibitemOpen
  \bibfield  {author} {\bibinfo {author} {\bibfnamefont {N.}~\bibnamefont
  {Yunes}}\ and\ \bibinfo {author} {\bibfnamefont {F.}~\bibnamefont
  {Pretorius}},\ }\bibfield  {title} {\bibinfo {title} {{Fundamental
  Theoretical Bias in Gravitational Wave Astrophysics and the Parameterized
  Post-Einsteinian Framework}},\ }\href
  {https://doi.org/10.1103/PhysRevD.80.122003} {\bibfield  {journal} {\bibinfo
  {journal} {Phys. Rev. D}\ }\textbf {\bibinfo {volume} {80}},\ \bibinfo
  {pages} {122003} (\bibinfo {year} {2009})},\ \Eprint
  {https://arxiv.org/abs/0909.3328} {arXiv:0909.3328 [gr-qc]} \BibitemShut
  {NoStop}%
\bibitem [{\citenamefont {{Li}}\ \emph {et~al.}(2012)\citenamefont {{Li}},
  \citenamefont {{Del Pozzo}}, \citenamefont {{Vitale}}, \citenamefont {{Van
  Den Broeck}}, \citenamefont {{Agathos}}, \citenamefont {{Veitch}},
  \citenamefont {{Grover}}, \citenamefont {{Sidery}}, \citenamefont
  {{Sturani}},\ and\ \citenamefont {{Vecchio}}}]{Li2012PN}%
  \BibitemOpen
  \bibfield  {author} {\bibinfo {author} {\bibfnamefont {T.~G.~F.}\
  \bibnamefont {{Li}}}, \bibinfo {author} {\bibfnamefont {W.}~\bibnamefont
  {{Del Pozzo}}}, \bibinfo {author} {\bibfnamefont {S.}~\bibnamefont
  {{Vitale}}}, \bibinfo {author} {\bibfnamefont {C.}~\bibnamefont {{Van Den
  Broeck}}}, \bibinfo {author} {\bibfnamefont {M.}~\bibnamefont {{Agathos}}},
  \bibinfo {author} {\bibfnamefont {J.}~\bibnamefont {{Veitch}}}, \bibinfo
  {author} {\bibfnamefont {K.}~\bibnamefont {{Grover}}}, \bibinfo {author}
  {\bibfnamefont {T.}~\bibnamefont {{Sidery}}}, \bibinfo {author}
  {\bibfnamefont {R.}~\bibnamefont {{Sturani}}},\ and\ \bibinfo {author}
  {\bibfnamefont {A.}~\bibnamefont {{Vecchio}}},\ }\bibfield  {title} {\bibinfo
  {title} {{Towards a generic test of the strong field dynamics of general
  relativity using compact binary coalescence}},\ }\href
  {https://doi.org/10.1103/PhysRevD.85.082003} {\bibfield  {journal} {\bibinfo
  {journal} {\prd}\ }\textbf {\bibinfo {volume} {85}},\ \bibinfo {eid} {082003}
  (\bibinfo {year} {2012})},\ \Eprint {https://arxiv.org/abs/1110.0530}
  {arXiv:1110.0530 [gr-qc]} \BibitemShut {NoStop}%
\bibitem [{\citenamefont {{Agathos}}\ \emph {et~al.}(2014)\citenamefont
  {{Agathos}}, \citenamefont {{Del Pozzo}}, \citenamefont {{Li}}, \citenamefont
  {{Van Den Broeck}}, \citenamefont {{Veitch}},\ and\ \citenamefont
  {{Vitale}}}]{agathos2014TIGER}%
  \BibitemOpen
  \bibfield  {author} {\bibinfo {author} {\bibfnamefont {M.}~\bibnamefont
  {{Agathos}}}, \bibinfo {author} {\bibfnamefont {W.}~\bibnamefont {{Del
  Pozzo}}}, \bibinfo {author} {\bibfnamefont {T.~G.~F.}\ \bibnamefont {{Li}}},
  \bibinfo {author} {\bibfnamefont {C.}~\bibnamefont {{Van Den Broeck}}},
  \bibinfo {author} {\bibfnamefont {J.}~\bibnamefont {{Veitch}}},\ and\
  \bibinfo {author} {\bibfnamefont {S.}~\bibnamefont {{Vitale}}},\ }\bibfield
  {title} {\bibinfo {title} {{TIGER: A data analysis pipeline for testing the
  strong-field dynamics of general relativity with gravitational wave signals
  from coalescing compact binaries}},\ }\href
  {https://doi.org/10.1103/PhysRevD.89.082001} {\bibfield  {journal} {\bibinfo
  {journal} {\prd}\ }\textbf {\bibinfo {volume} {89}},\ \bibinfo {eid} {082001}
  (\bibinfo {year} {2014})},\ \Eprint {https://arxiv.org/abs/1311.0420}
  {arXiv:1311.0420 [gr-qc]} \BibitemShut {NoStop}%
\bibitem [{\citenamefont {Mehta}\ \emph {et~al.}(2023)\citenamefont {Mehta},
  \citenamefont {Buonanno}, \citenamefont {Cotesta}, \citenamefont {Ghosh},
  \citenamefont {Sennett},\ and\ \citenamefont {Steinhoff}}]{Mehta:2022pcn}%
  \BibitemOpen
  \bibfield  {author} {\bibinfo {author} {\bibfnamefont {A.~K.}\ \bibnamefont
  {Mehta}}, \bibinfo {author} {\bibfnamefont {A.}~\bibnamefont {Buonanno}},
  \bibinfo {author} {\bibfnamefont {R.}~\bibnamefont {Cotesta}}, \bibinfo
  {author} {\bibfnamefont {A.}~\bibnamefont {Ghosh}}, \bibinfo {author}
  {\bibfnamefont {N.}~\bibnamefont {Sennett}},\ and\ \bibinfo {author}
  {\bibfnamefont {J.}~\bibnamefont {Steinhoff}},\ }\bibfield  {title} {\bibinfo
  {title} {{Tests of general relativity with gravitational-wave observations
  using a flexible theory-independent method}},\ }\href
  {https://doi.org/10.1103/PhysRevD.107.044020} {\bibfield  {journal} {\bibinfo
   {journal} {Phys. Rev. D}\ }\textbf {\bibinfo {volume} {107}},\ \bibinfo
  {pages} {044020} (\bibinfo {year} {2023})},\ \Eprint
  {https://arxiv.org/abs/2203.13937} {arXiv:2203.13937 [gr-qc]} \BibitemShut
  {NoStop}%
\bibitem [{\citenamefont {Carullo}\ \emph {et~al.}(2019)\citenamefont
  {Carullo}, \citenamefont {Del~Pozzo},\ and\ \citenamefont
  {Veitch}}]{Carullo:2019flw}%
  \BibitemOpen
  \bibfield  {author} {\bibinfo {author} {\bibfnamefont {G.}~\bibnamefont
  {Carullo}}, \bibinfo {author} {\bibfnamefont {W.}~\bibnamefont {Del~Pozzo}},\
  and\ \bibinfo {author} {\bibfnamefont {J.}~\bibnamefont {Veitch}},\
  }\bibfield  {title} {\bibinfo {title} {{Observational Black Hole
  Spectroscopy: A time-domain multimode analysis of GW150914}},\ }\href
  {https://doi.org/10.1103/PhysRevD.99.123029} {\bibfield  {journal} {\bibinfo
  {journal} {Phys. Rev. D}\ }\textbf {\bibinfo {volume} {99}},\ \bibinfo
  {pages} {123029} (\bibinfo {year} {2019})},\ \bibinfo {note} {[Erratum:
  Phys.Rev.D 100, 089903 (2019)]},\ \Eprint {https://arxiv.org/abs/1902.07527}
  {arXiv:1902.07527 [gr-qc]} \BibitemShut {NoStop}%
\bibitem [{\citenamefont {Isi}\ \emph {et~al.}(2019{\natexlab{b}})\citenamefont
  {Isi}, \citenamefont {Giesler}, \citenamefont {Farr}, \citenamefont
  {Scheel},\ and\ \citenamefont {Teukolsky}}]{Isi:2019aib}%
  \BibitemOpen
  \bibfield  {author} {\bibinfo {author} {\bibfnamefont {M.}~\bibnamefont
  {Isi}}, \bibinfo {author} {\bibfnamefont {M.}~\bibnamefont {Giesler}},
  \bibinfo {author} {\bibfnamefont {W.~M.}\ \bibnamefont {Farr}}, \bibinfo
  {author} {\bibfnamefont {M.~A.}\ \bibnamefont {Scheel}},\ and\ \bibinfo
  {author} {\bibfnamefont {S.~A.}\ \bibnamefont {Teukolsky}},\ }\bibfield
  {title} {\bibinfo {title} {{Testing the no-hair theorem with GW150914}},\
  }\href {https://doi.org/10.1103/PhysRevLett.123.111102} {\bibfield  {journal}
  {\bibinfo  {journal} {Phys. Rev. Lett.}\ }\textbf {\bibinfo {volume} {123}},\
  \bibinfo {pages} {111102} (\bibinfo {year} {2019}{\natexlab{b}})},\ \Eprint
  {https://arxiv.org/abs/1905.00869} {arXiv:1905.00869 [gr-qc]} \BibitemShut
  {NoStop}%
\bibitem [{\citenamefont {Ghosh}\ \emph
  {et~al.}(2016{\natexlab{a}})\citenamefont {Ghosh} \emph
  {et~al.}}]{Ghosh2016qgn}%
  \BibitemOpen
  \bibfield  {author} {\bibinfo {author} {\bibfnamefont {A.}~\bibnamefont
  {Ghosh}} \emph {et~al.},\ }\bibfield  {title} {\bibinfo {title} {{Testing
  general relativity using golden black-hole binaries}},\ }\href
  {https://doi.org/10.1103/PhysRevD.94.021101} {\bibfield  {journal} {\bibinfo
  {journal} {Phys. Rev. D}\ }\textbf {\bibinfo {volume} {94}},\ \bibinfo
  {pages} {021101} (\bibinfo {year} {2016}{\natexlab{a}})},\ \Eprint
  {https://arxiv.org/abs/1602.02453} {arXiv:1602.02453 [gr-qc]} \BibitemShut
  {NoStop}%
\bibitem [{\citenamefont {Ghosh}\ \emph
  {et~al.}(2018{\natexlab{a}})\citenamefont {Ghosh}, \citenamefont
  {Johnson-Mcdaniel}, \citenamefont {Ghosh}, \citenamefont {Mishra},
  \citenamefont {Ajith}, \citenamefont {Del~Pozzo}, \citenamefont {Berry},
  \citenamefont {Nielsen},\ and\ \citenamefont {London}}]{Ghosh2017gfp}%
  \BibitemOpen
  \bibfield  {author} {\bibinfo {author} {\bibfnamefont {A.}~\bibnamefont
  {Ghosh}}, \bibinfo {author} {\bibfnamefont {N.~K.}\ \bibnamefont
  {Johnson-Mcdaniel}}, \bibinfo {author} {\bibfnamefont {A.}~\bibnamefont
  {Ghosh}}, \bibinfo {author} {\bibfnamefont {C.~K.}\ \bibnamefont {Mishra}},
  \bibinfo {author} {\bibfnamefont {P.}~\bibnamefont {Ajith}}, \bibinfo
  {author} {\bibfnamefont {W.}~\bibnamefont {Del~Pozzo}}, \bibinfo {author}
  {\bibfnamefont {C.~P.~L.}\ \bibnamefont {Berry}}, \bibinfo {author}
  {\bibfnamefont {A.~B.}\ \bibnamefont {Nielsen}},\ and\ \bibinfo {author}
  {\bibfnamefont {L.}~\bibnamefont {London}},\ }\bibfield  {title} {\bibinfo
  {title} {{Testing general relativity using gravitational wave signals from
  the inspiral, merger and ringdown of binary black holes}},\ }\href
  {https://doi.org/10.1088/1361-6382/aa972e} {\bibfield  {journal} {\bibinfo
  {journal} {Class. Quant. Grav.}\ }\textbf {\bibinfo {volume} {35}},\ \bibinfo
  {pages} {014002} (\bibinfo {year} {2018}{\natexlab{a}})},\ \Eprint
  {https://arxiv.org/abs/1704.06784} {arXiv:1704.06784 [gr-qc]} \BibitemShut
  {NoStop}%
\bibitem [{\citenamefont {Ng}\ \emph {et~al.}(2023)\citenamefont {Ng},
  \citenamefont {Isi}, \citenamefont {Wong},\ and\ \citenamefont
  {Farr}}]{Ng:2023jjt}%
  \BibitemOpen
  \bibfield  {author} {\bibinfo {author} {\bibfnamefont {T.~C.~K.}\
  \bibnamefont {Ng}}, \bibinfo {author} {\bibfnamefont {M.}~\bibnamefont
  {Isi}}, \bibinfo {author} {\bibfnamefont {K.~W.~K.}\ \bibnamefont {Wong}},\
  and\ \bibinfo {author} {\bibfnamefont {W.~M.}\ \bibnamefont {Farr}},\
  }\bibfield  {title} {\bibinfo {title} {{Constraining gravitational wave
  amplitude birefringence with GWTC-3}},\ }\href
  {https://doi.org/10.1103/PhysRevD.108.084068} {\bibfield  {journal} {\bibinfo
   {journal} {Phys. Rev. D}\ }\textbf {\bibinfo {volume} {108}},\ \bibinfo
  {pages} {084068} (\bibinfo {year} {2023})},\ \Eprint
  {https://arxiv.org/abs/2305.05844} {arXiv:2305.05844 [gr-qc]} \BibitemShut
  {NoStop}%
\bibitem [{\citenamefont {Abbott}\ \emph {et~al.}(2016)\citenamefont {Abbott}
  \emph {et~al.}}]{LIGOScientific:2016lio}%
  \BibitemOpen
  \bibfield  {author} {\bibinfo {author} {\bibfnamefont {B.~P.}\ \bibnamefont
  {Abbott}} \emph {et~al.} (\bibinfo {collaboration} {LIGO Scientific,
  Virgo}),\ }\bibfield  {title} {\bibinfo {title} {{Tests of general relativity
  with GW150914}},\ }\href {https://doi.org/10.1103/PhysRevLett.116.221101}
  {\bibfield  {journal} {\bibinfo  {journal} {Phys. Rev. Lett.}\ }\textbf
  {\bibinfo {volume} {116}},\ \bibinfo {pages} {221101} (\bibinfo {year}
  {2016})},\ \bibinfo {note} {[Erratum: Phys.Rev.Lett. 121, 129902 (2018)]},\
  \Eprint {https://arxiv.org/abs/1602.03841} {arXiv:1602.03841 [gr-qc]}
  \BibitemShut {NoStop}%
\bibitem [{\citenamefont {Perkins}\ and\ \citenamefont
  {Yunes}(2022)}]{Perkins:2022fhr}%
  \BibitemOpen
  \bibfield  {author} {\bibinfo {author} {\bibfnamefont {S.}~\bibnamefont
  {Perkins}}\ and\ \bibinfo {author} {\bibfnamefont {N.}~\bibnamefont
  {Yunes}},\ }\bibfield  {title} {\bibinfo {title} {{Are parametrized tests of
  general relativity with gravitational waves robust to unknown higher
  post-Newtonian order effects?}},\ }\href
  {https://doi.org/10.1103/PhysRevD.105.124047} {\bibfield  {journal} {\bibinfo
   {journal} {Phys. Rev. D}\ }\textbf {\bibinfo {volume} {105}},\ \bibinfo
  {pages} {124047} (\bibinfo {year} {2022})},\ \Eprint
  {https://arxiv.org/abs/2201.02542} {arXiv:2201.02542 [gr-qc]} \BibitemShut
  {NoStop}%
\bibitem [{\citenamefont {Pai}\ and\ \citenamefont {Arun}(2013)}]{Pai:2012mv}%
  \BibitemOpen
  \bibfield  {author} {\bibinfo {author} {\bibfnamefont {A.}~\bibnamefont
  {Pai}}\ and\ \bibinfo {author} {\bibfnamefont {K.~G.}\ \bibnamefont {Arun}},\
  }\bibfield  {title} {\bibinfo {title} {{Singular value decomposition in
  parametrised tests of post-Newtonian theory}},\ }\href
  {https://doi.org/10.1088/0264-9381/30/2/025011} {\bibfield  {journal}
  {\bibinfo  {journal} {Class. Quant. Grav.}\ }\textbf {\bibinfo {volume}
  {30}},\ \bibinfo {pages} {025011} (\bibinfo {year} {2013})},\ \Eprint
  {https://arxiv.org/abs/1207.1943} {arXiv:1207.1943 [gr-qc]} \BibitemShut
  {NoStop}%
\bibitem [{\citenamefont {Saleem}\ \emph {et~al.}(2022)\citenamefont {Saleem},
  \citenamefont {Datta}, \citenamefont {Arun},\ and\ \citenamefont
  {Sathyaprakash}}]{Saleem:2021nsb}%
  \BibitemOpen
  \bibfield  {author} {\bibinfo {author} {\bibfnamefont {M.}~\bibnamefont
  {Saleem}}, \bibinfo {author} {\bibfnamefont {S.}~\bibnamefont {Datta}},
  \bibinfo {author} {\bibfnamefont {K.~G.}\ \bibnamefont {Arun}},\ and\
  \bibinfo {author} {\bibfnamefont {B.~S.}\ \bibnamefont {Sathyaprakash}},\
  }\bibfield  {title} {\bibinfo {title} {{Parametrized tests of post-Newtonian
  theory using principal component analysis}},\ }\href
  {https://doi.org/10.1103/PhysRevD.105.084062} {\bibfield  {journal} {\bibinfo
   {journal} {Phys. Rev. D}\ }\textbf {\bibinfo {volume} {105}},\ \bibinfo
  {pages} {084062} (\bibinfo {year} {2022})},\ \Eprint
  {https://arxiv.org/abs/2110.10147} {arXiv:2110.10147 [gr-qc]} \BibitemShut
  {NoStop}%
\bibitem [{\citenamefont {Lewandowski}\ \emph {et~al.}(2009)\citenamefont
  {Lewandowski}, \citenamefont {Kurowicka},\ and\ \citenamefont {Joe}}]{LKJ}%
  \BibitemOpen
  \bibfield  {author} {\bibinfo {author} {\bibfnamefont {D.}~\bibnamefont
  {Lewandowski}}, \bibinfo {author} {\bibfnamefont {D.}~\bibnamefont
  {Kurowicka}},\ and\ \bibinfo {author} {\bibfnamefont {H.}~\bibnamefont
  {Joe}},\ }\bibfield  {title} {\bibinfo {title} {Generating random correlation
  matrices based on vines and extended onion method},\ }\href
  {https://doi.org/https://doi.org/10.1016/j.jmva.2009.04.008} {\bibfield
  {journal} {\bibinfo  {journal} {Journal of Multivariate Analysis}\ }\textbf
  {\bibinfo {volume} {100}},\ \bibinfo {pages} {1989} (\bibinfo {year}
  {2009})}\BibitemShut {NoStop}%
\bibitem [{\citenamefont {Akinc}\ and\ \citenamefont
  {Vandebroek}(2018)}]{Akinc_2018}%
  \BibitemOpen
  \bibfield  {author} {\bibinfo {author} {\bibfnamefont {D.}~\bibnamefont
  {Akinc}}\ and\ \bibinfo {author} {\bibfnamefont {M.}~\bibnamefont
  {Vandebroek}},\ }\bibfield  {title} {\bibinfo {title} {Bayesian estimation of
  mixed logit models: Selecting an appropriate prior for the covariance
  matrix},\ }\href {https://doi.org/https://doi.org/10.1016/j.jocm.2017.11.004}
  {\bibfield  {journal} {\bibinfo  {journal} {Journal of Choice Modelling}\
  }\textbf {\bibinfo {volume} {29}},\ \bibinfo {pages} {133} (\bibinfo {year}
  {2018})}\BibitemShut {NoStop}%
\bibitem [{\citenamefont {Lieu}\ \emph {et~al.}(2017)\citenamefont {Lieu},
  \citenamefont {Farr}, \citenamefont {Betancourt}, \citenamefont {Smith},
  \citenamefont {Sereno},\ and\ \citenamefont {McCarthy}}]{Liue_2017}%
  \BibitemOpen
  \bibfield  {author} {\bibinfo {author} {\bibfnamefont {M.}~\bibnamefont
  {Lieu}}, \bibinfo {author} {\bibfnamefont {W.~M.}\ \bibnamefont {Farr}},
  \bibinfo {author} {\bibfnamefont {M.}~\bibnamefont {Betancourt}}, \bibinfo
  {author} {\bibfnamefont {G.~P.}\ \bibnamefont {Smith}}, \bibinfo {author}
  {\bibfnamefont {M.}~\bibnamefont {Sereno}},\ and\ \bibinfo {author}
  {\bibfnamefont {I.~G.}\ \bibnamefont {McCarthy}},\ }\bibfield  {title}
  {\bibinfo {title} {{Hierarchical inference of the relationship between
  concentration and mass in galaxy groups and clusters}},\ }\href
  {https://doi.org/10.1093/mnras/stx686} {\bibfield  {journal} {\bibinfo
  {journal} {Monthly Notices of the Royal Astronomical Society}\ }\textbf
  {\bibinfo {volume} {468}},\ \bibinfo {pages} {4872} (\bibinfo {year}
  {2017})},\ \Eprint
  {https://arxiv.org/abs/https://academic.oup.com/mnras/article-pdf/468/4/4872/16638371/stx686.pdf}
  {https://academic.oup.com/mnras/article-pdf/468/4/4872/16638371/stx686.pdf}
  \BibitemShut {NoStop}%
\bibitem [{\citenamefont {Tao}\ \emph {et~al.}(0)\citenamefont {Tao},
  \citenamefont {Phoon}, \citenamefont {Sun},\ and\ \citenamefont
  {Cai}}]{Tao_2022}%
  \BibitemOpen
  \bibfield  {author} {\bibinfo {author} {\bibfnamefont {Y.}~\bibnamefont
  {Tao}}, \bibinfo {author} {\bibfnamefont {K.-K.}\ \bibnamefont {Phoon}},
  \bibinfo {author} {\bibfnamefont {H.}~\bibnamefont {Sun}},\ and\ \bibinfo
  {author} {\bibfnamefont {Y.}~\bibnamefont {Cai}},\ }\bibfield  {title}
  {\bibinfo {title} {Hierarchical bayesian model for predicting small-strain
  stiffness of sand},\ }\href {https://doi.org/10.1139/cgj-2022-0598}
  {\bibfield  {journal} {\bibinfo  {journal} {Canadian Geotechnical Journal}\
  }\textbf {\bibinfo {volume} {0}},\ \bibinfo {pages} {null} (\bibinfo {year}
  {0})},\ \Eprint {https://arxiv.org/abs/https://doi.org/10.1139/cgj-2022-0598}
  {https://doi.org/10.1139/cgj-2022-0598} \BibitemShut {NoStop}%
\bibitem [{\citenamefont {Feng}\ \emph {et~al.}(2021)\citenamefont {Feng},
  \citenamefont {Gao}, \citenamefont {Mignan},\ and\ \citenamefont
  {Li}}]{Feng_2021}%
  \BibitemOpen
  \bibfield  {author} {\bibinfo {author} {\bibfnamefont {Y.}~\bibnamefont
  {Feng}}, \bibinfo {author} {\bibfnamefont {K.}~\bibnamefont {Gao}}, \bibinfo
  {author} {\bibfnamefont {A.}~\bibnamefont {Mignan}},\ and\ \bibinfo {author}
  {\bibfnamefont {J.}~\bibnamefont {Li}},\ }\bibfield  {title} {\bibinfo
  {title} {Improving local mean stress estimation using bayesian hierarchical
  modelling},\ }\href
  {https://doi.org/https://doi.org/10.1016/j.ijrmms.2021.104924} {\bibfield
  {journal} {\bibinfo  {journal} {International Journal of Rock Mechanics and
  Mining Sciences}\ }\textbf {\bibinfo {volume} {148}},\ \bibinfo {pages}
  {104924} (\bibinfo {year} {2021})}\BibitemShut {NoStop}%
\bibitem [{\citenamefont {Golomb}\ and\ \citenamefont
  {Talbot}(2022)}]{Golomb:2021tll}%
  \BibitemOpen
  \bibfield  {author} {\bibinfo {author} {\bibfnamefont {J.}~\bibnamefont
  {Golomb}}\ and\ \bibinfo {author} {\bibfnamefont {C.}~\bibnamefont
  {Talbot}},\ }\bibfield  {title} {\bibinfo {title} {{Hierarchical Inference of
  Binary Neutron Star Mass Distribution and Equation of State with
  Gravitational Waves}},\ }\href {https://doi.org/10.3847/1538-4357/ac43bc}
  {\bibfield  {journal} {\bibinfo  {journal} {Astrophys. J.}\ }\textbf
  {\bibinfo {volume} {926}},\ \bibinfo {pages} {79} (\bibinfo {year} {2022})},\
  \Eprint {https://arxiv.org/abs/2106.15745} {arXiv:2106.15745 [astro-ph.HE]}
  \BibitemShut {NoStop}%
\bibitem [{\citenamefont {Ghosh}\ \emph
  {et~al.}(2016{\natexlab{b}})\citenamefont {Ghosh} \emph
  {et~al.}}]{Ghosh:2016qgn}%
  \BibitemOpen
  \bibfield  {author} {\bibinfo {author} {\bibfnamefont {A.}~\bibnamefont
  {Ghosh}} \emph {et~al.},\ }\bibfield  {title} {\bibinfo {title} {{Testing
  general relativity using golden black-hole binaries}},\ }\href
  {https://doi.org/10.1103/PhysRevD.94.021101} {\bibfield  {journal} {\bibinfo
  {journal} {Phys. Rev. D}\ }\textbf {\bibinfo {volume} {94}},\ \bibinfo
  {pages} {021101} (\bibinfo {year} {2016}{\natexlab{b}})},\ \Eprint
  {https://arxiv.org/abs/1602.02453} {arXiv:1602.02453 [gr-qc]} \BibitemShut
  {NoStop}%
\bibitem [{\citenamefont {Ghosh}\ \emph
  {et~al.}(2018{\natexlab{b}})\citenamefont {Ghosh}, \citenamefont
  {Johnson-Mcdaniel}, \citenamefont {Ghosh}, \citenamefont {Mishra},
  \citenamefont {Ajith}, \citenamefont {Del~Pozzo}, \citenamefont {Berry},
  \citenamefont {Nielsen},\ and\ \citenamefont {London}}]{Ghosh:2017gfp}%
  \BibitemOpen
  \bibfield  {author} {\bibinfo {author} {\bibfnamefont {A.}~\bibnamefont
  {Ghosh}}, \bibinfo {author} {\bibfnamefont {N.~K.}\ \bibnamefont
  {Johnson-Mcdaniel}}, \bibinfo {author} {\bibfnamefont {A.}~\bibnamefont
  {Ghosh}}, \bibinfo {author} {\bibfnamefont {C.~K.}\ \bibnamefont {Mishra}},
  \bibinfo {author} {\bibfnamefont {P.}~\bibnamefont {Ajith}}, \bibinfo
  {author} {\bibfnamefont {W.}~\bibnamefont {Del~Pozzo}}, \bibinfo {author}
  {\bibfnamefont {C.~P.~L.}\ \bibnamefont {Berry}}, \bibinfo {author}
  {\bibfnamefont {A.~B.}\ \bibnamefont {Nielsen}},\ and\ \bibinfo {author}
  {\bibfnamefont {L.}~\bibnamefont {London}},\ }\bibfield  {title} {\bibinfo
  {title} {{Testing general relativity using gravitational wave signals from
  the inspiral, merger and ringdown of binary black holes}},\ }\href
  {https://doi.org/10.1088/1361-6382/aa972e} {\bibfield  {journal} {\bibinfo
  {journal} {Class. Quant. Grav.}\ }\textbf {\bibinfo {volume} {35}},\ \bibinfo
  {pages} {014002} (\bibinfo {year} {2018}{\natexlab{b}})},\ \Eprint
  {https://arxiv.org/abs/1704.06784} {arXiv:1704.06784 [gr-qc]} \BibitemShut
  {NoStop}%
\bibitem [{\citenamefont {Isi}\ \emph {et~al.}(2021)\citenamefont {Isi},
  \citenamefont {Farr}, \citenamefont {Giesler}, \citenamefont {Scheel},\ and\
  \citenamefont {Teukolsky}}]{Isi:2020tac}%
  \BibitemOpen
  \bibfield  {author} {\bibinfo {author} {\bibfnamefont {M.}~\bibnamefont
  {Isi}}, \bibinfo {author} {\bibfnamefont {W.~M.}\ \bibnamefont {Farr}},
  \bibinfo {author} {\bibfnamefont {M.}~\bibnamefont {Giesler}}, \bibinfo
  {author} {\bibfnamefont {M.~A.}\ \bibnamefont {Scheel}},\ and\ \bibinfo
  {author} {\bibfnamefont {S.~A.}\ \bibnamefont {Teukolsky}},\ }\bibfield
  {title} {\bibinfo {title} {{Testing the Black-Hole Area Law with GW150914}},\
  }\href {https://doi.org/10.1103/PhysRevLett.127.011103} {\bibfield  {journal}
  {\bibinfo  {journal} {Phys. Rev. Lett.}\ }\textbf {\bibinfo {volume} {127}},\
  \bibinfo {pages} {011103} (\bibinfo {year} {2021})},\ \Eprint
  {https://arxiv.org/abs/2012.04486} {arXiv:2012.04486 [gr-qc]} \BibitemShut
  {NoStop}%
\bibitem [{\citenamefont {Cabero}\ \emph {et~al.}(2018)\citenamefont {Cabero},
  \citenamefont {Capano}, \citenamefont {Fischer-Birnholtz}, \citenamefont
  {Krishnan}, \citenamefont {Nielsen}, \citenamefont {Nitz},\ and\
  \citenamefont {Biwer}}]{Cabero:2017avf}%
  \BibitemOpen
  \bibfield  {author} {\bibinfo {author} {\bibfnamefont {M.}~\bibnamefont
  {Cabero}}, \bibinfo {author} {\bibfnamefont {C.~D.}\ \bibnamefont {Capano}},
  \bibinfo {author} {\bibfnamefont {O.}~\bibnamefont {Fischer-Birnholtz}},
  \bibinfo {author} {\bibfnamefont {B.}~\bibnamefont {Krishnan}}, \bibinfo
  {author} {\bibfnamefont {A.~B.}\ \bibnamefont {Nielsen}}, \bibinfo {author}
  {\bibfnamefont {A.~H.}\ \bibnamefont {Nitz}},\ and\ \bibinfo {author}
  {\bibfnamefont {C.~M.}\ \bibnamefont {Biwer}},\ }\bibfield  {title} {\bibinfo
  {title} {{Observational tests of the black hole area increase law}},\ }\href
  {https://doi.org/10.1103/PhysRevD.97.124069} {\bibfield  {journal} {\bibinfo
  {journal} {Phys. Rev. D}\ }\textbf {\bibinfo {volume} {97}},\ \bibinfo
  {pages} {124069} (\bibinfo {year} {2018})},\ \Eprint
  {https://arxiv.org/abs/1711.09073} {arXiv:1711.09073 [gr-qc]} \BibitemShut
  {NoStop}%
\bibitem [{\citenamefont {Schmidt}\ \emph {et~al.}(2011)\citenamefont
  {Schmidt}, \citenamefont {Hannam}, \citenamefont {Husa},\ and\ \citenamefont
  {Ajith}}]{Schmidt:2010it}%
  \BibitemOpen
  \bibfield  {author} {\bibinfo {author} {\bibfnamefont {P.}~\bibnamefont
  {Schmidt}}, \bibinfo {author} {\bibfnamefont {M.}~\bibnamefont {Hannam}},
  \bibinfo {author} {\bibfnamefont {S.}~\bibnamefont {Husa}},\ and\ \bibinfo
  {author} {\bibfnamefont {P.}~\bibnamefont {Ajith}},\ }\bibfield  {title}
  {\bibinfo {title} {{Tracking the precession of compact binaries from their
  gravitational-wave signal}},\ }\href
  {https://doi.org/10.1103/PhysRevD.84.024046} {\bibfield  {journal} {\bibinfo
  {journal} {Phys. Rev. D}\ }\textbf {\bibinfo {volume} {84}},\ \bibinfo
  {pages} {024046} (\bibinfo {year} {2011})},\ \Eprint
  {https://arxiv.org/abs/1012.2879} {arXiv:1012.2879 [gr-qc]} \BibitemShut
  {NoStop}%
\bibitem [{\citenamefont {Schmidt}\ \emph {et~al.}(2012)\citenamefont
  {Schmidt}, \citenamefont {Hannam},\ and\ \citenamefont
  {Husa}}]{Schmidt:2012rh}%
  \BibitemOpen
  \bibfield  {author} {\bibinfo {author} {\bibfnamefont {P.}~\bibnamefont
  {Schmidt}}, \bibinfo {author} {\bibfnamefont {M.}~\bibnamefont {Hannam}},\
  and\ \bibinfo {author} {\bibfnamefont {S.}~\bibnamefont {Husa}},\ }\bibfield
  {title} {\bibinfo {title} {{Towards models of gravitational waveforms from
  generic binaries: A simple approximate mapping between precessing and
  non-precessing inspiral signals}},\ }\href
  {https://doi.org/10.1103/PhysRevD.86.104063} {\bibfield  {journal} {\bibinfo
  {journal} {Phys. Rev. D}\ }\textbf {\bibinfo {volume} {86}},\ \bibinfo
  {pages} {104063} (\bibinfo {year} {2012})},\ \Eprint
  {https://arxiv.org/abs/1207.3088} {arXiv:1207.3088 [gr-qc]} \BibitemShut
  {NoStop}%
\bibitem [{\citenamefont {Hannam}\ \emph {et~al.}(2014)\citenamefont {Hannam},
  \citenamefont {Schmidt}, \citenamefont {Boh\'e}, \citenamefont {Haegel},
  \citenamefont {Husa}, \citenamefont {Ohme}, \citenamefont {Pratten},\ and\
  \citenamefont {P\"urrer}}]{Hannam:2013oca}%
  \BibitemOpen
  \bibfield  {author} {\bibinfo {author} {\bibfnamefont {M.}~\bibnamefont
  {Hannam}}, \bibinfo {author} {\bibfnamefont {P.}~\bibnamefont {Schmidt}},
  \bibinfo {author} {\bibfnamefont {A.}~\bibnamefont {Boh\'e}}, \bibinfo
  {author} {\bibfnamefont {L.}~\bibnamefont {Haegel}}, \bibinfo {author}
  {\bibfnamefont {S.}~\bibnamefont {Husa}}, \bibinfo {author} {\bibfnamefont
  {F.}~\bibnamefont {Ohme}}, \bibinfo {author} {\bibfnamefont {G.}~\bibnamefont
  {Pratten}},\ and\ \bibinfo {author} {\bibfnamefont {M.}~\bibnamefont
  {P\"urrer}},\ }\bibfield  {title} {\bibinfo {title} {{Simple Model of
  Complete Precessing Black-Hole-Binary Gravitational Waveforms}},\ }\href
  {https://doi.org/10.1103/PhysRevLett.113.151101} {\bibfield  {journal}
  {\bibinfo  {journal} {Phys. Rev. Lett.}\ }\textbf {\bibinfo {volume} {113}},\
  \bibinfo {pages} {151101} (\bibinfo {year} {2014})},\ \Eprint
  {https://arxiv.org/abs/1308.3271} {arXiv:1308.3271 [gr-qc]} \BibitemShut
  {NoStop}%
\bibitem [{\citenamefont {Khan}\ \emph {et~al.}(2020)\citenamefont {Khan},
  \citenamefont {Ohme}, \citenamefont {Chatziioannou},\ and\ \citenamefont
  {Hannam}}]{Khan:2019kot}%
  \BibitemOpen
  \bibfield  {author} {\bibinfo {author} {\bibfnamefont {S.}~\bibnamefont
  {Khan}}, \bibinfo {author} {\bibfnamefont {F.}~\bibnamefont {Ohme}}, \bibinfo
  {author} {\bibfnamefont {K.}~\bibnamefont {Chatziioannou}},\ and\ \bibinfo
  {author} {\bibfnamefont {M.}~\bibnamefont {Hannam}},\ }\bibfield  {title}
  {\bibinfo {title} {{Including higher order multipoles in gravitational-wave
  models for precessing binary black holes}},\ }\href
  {https://doi.org/10.1103/PhysRevD.101.024056} {\bibfield  {journal} {\bibinfo
   {journal} {Phys. Rev. D}\ }\textbf {\bibinfo {volume} {101}},\ \bibinfo
  {pages} {024056} (\bibinfo {year} {2020})},\ \Eprint
  {https://arxiv.org/abs/1911.06050} {arXiv:1911.06050 [gr-qc]} \BibitemShut
  {NoStop}%
\bibitem [{\citenamefont {Pratten}\ \emph {et~al.}(2021)\citenamefont {Pratten}
  \emph {et~al.}}]{Pratten:2020ceb}%
  \BibitemOpen
  \bibfield  {author} {\bibinfo {author} {\bibfnamefont {G.}~\bibnamefont
  {Pratten}} \emph {et~al.},\ }\bibfield  {title} {\bibinfo {title}
  {{Computationally efficient models for the dominant and subdominant harmonic
  modes of precessing binary black holes}},\ }\href
  {https://doi.org/10.1103/PhysRevD.103.104056} {\bibfield  {journal} {\bibinfo
   {journal} {Phys. Rev. D}\ }\textbf {\bibinfo {volume} {103}},\ \bibinfo
  {pages} {104056} (\bibinfo {year} {2021})},\ \Eprint
  {https://arxiv.org/abs/2004.06503} {arXiv:2004.06503 [gr-qc]} \BibitemShut
  {NoStop}%
\bibitem [{\citenamefont {Garc\'\i{}a-Quir\'os}\ \emph
  {et~al.}(2020)\citenamefont {Garc\'\i{}a-Quir\'os}, \citenamefont {Colleoni},
  \citenamefont {Husa}, \citenamefont {Estell\'es}, \citenamefont {Pratten},
  \citenamefont {Ramos-Buades}, \citenamefont {Mateu-Lucena},\ and\
  \citenamefont {Jaume}}]{Garcia-Quiros:2020qpx}%
  \BibitemOpen
  \bibfield  {author} {\bibinfo {author} {\bibfnamefont {C.}~\bibnamefont
  {Garc\'\i{}a-Quir\'os}}, \bibinfo {author} {\bibfnamefont {M.}~\bibnamefont
  {Colleoni}}, \bibinfo {author} {\bibfnamefont {S.}~\bibnamefont {Husa}},
  \bibinfo {author} {\bibfnamefont {H.}~\bibnamefont {Estell\'es}}, \bibinfo
  {author} {\bibfnamefont {G.}~\bibnamefont {Pratten}}, \bibinfo {author}
  {\bibfnamefont {A.}~\bibnamefont {Ramos-Buades}}, \bibinfo {author}
  {\bibfnamefont {M.}~\bibnamefont {Mateu-Lucena}},\ and\ \bibinfo {author}
  {\bibfnamefont {R.}~\bibnamefont {Jaume}},\ }\bibfield  {title} {\bibinfo
  {title} {{Multimode frequency-domain model for the gravitational wave signal
  from nonprecessing black-hole binaries}},\ }\href
  {https://doi.org/10.1103/PhysRevD.102.064002} {\bibfield  {journal} {\bibinfo
   {journal} {Phys. Rev. D}\ }\textbf {\bibinfo {volume} {102}},\ \bibinfo
  {pages} {064002} (\bibinfo {year} {2020})},\ \Eprint
  {https://arxiv.org/abs/2001.10914} {arXiv:2001.10914 [gr-qc]} \BibitemShut
  {NoStop}%
\bibitem [{\citenamefont {Pratten}\ \emph {et~al.}(2020)\citenamefont
  {Pratten}, \citenamefont {Husa}, \citenamefont {Garcia-Quiros}, \citenamefont
  {Colleoni}, \citenamefont {Ramos-Buades}, \citenamefont {Estelles},\ and\
  \citenamefont {Jaume}}]{Pratten:2020fqn}%
  \BibitemOpen
  \bibfield  {author} {\bibinfo {author} {\bibfnamefont {G.}~\bibnamefont
  {Pratten}}, \bibinfo {author} {\bibfnamefont {S.}~\bibnamefont {Husa}},
  \bibinfo {author} {\bibfnamefont {C.}~\bibnamefont {Garcia-Quiros}}, \bibinfo
  {author} {\bibfnamefont {M.}~\bibnamefont {Colleoni}}, \bibinfo {author}
  {\bibfnamefont {A.}~\bibnamefont {Ramos-Buades}}, \bibinfo {author}
  {\bibfnamefont {H.}~\bibnamefont {Estelles}},\ and\ \bibinfo {author}
  {\bibfnamefont {R.}~\bibnamefont {Jaume}},\ }\bibfield  {title} {\bibinfo
  {title} {{Setting the cornerstone for a family of models for gravitational
  waves from compact binaries: The dominant harmonic for nonprecessing
  quasicircular black holes}},\ }\href
  {https://doi.org/10.1103/PhysRevD.102.064001} {\bibfield  {journal} {\bibinfo
   {journal} {Phys. Rev. D}\ }\textbf {\bibinfo {volume} {102}},\ \bibinfo
  {pages} {064001} (\bibinfo {year} {2020})},\ \Eprint
  {https://arxiv.org/abs/2001.11412} {arXiv:2001.11412 [gr-qc]} \BibitemShut
  {NoStop}%
\bibitem [{\citenamefont {Abbott}\ \emph {et~al.}(2020)\citenamefont {Abbott},
  \citenamefont {Abbott}, \citenamefont {Abraham} \emph {et~al.}}]{GW190814}%
  \BibitemOpen
  \bibfield  {author} {\bibinfo {author} {\bibfnamefont {R.}~\bibnamefont
  {Abbott}}, \bibinfo {author} {\bibfnamefont {T.~D.}\ \bibnamefont {Abbott}},
  \bibinfo {author} {\bibfnamefont {S.}~\bibnamefont {Abraham}}, \emph
  {et~al.},\ }\bibfield  {title} {\bibinfo {title} {Gw190814: Gravitational
  waves from the coalescence of a 23 solar mass black hole with a 2.6 solar
  mass compact object},\ }\href {https://doi.org/10.3847/2041-8213/ab960f}
  {\bibfield  {journal} {\bibinfo  {journal} {The Astrophysical Journal
  Letters}\ }\textbf {\bibinfo {volume} {896}},\ \bibinfo {pages} {L44}
  (\bibinfo {year} {2020})}\BibitemShut {NoStop}%
\bibitem [{\citenamefont {LIGO~Scientific}(2022)}]{data}%
  \BibitemOpen
  \bibfield  {author} {\bibinfo {author} {\bibfnamefont {K.}~\bibnamefont
  {LIGO~Scientific}, \bibfnamefont {VIRGO}},\ }\bibfield  {title} {\bibinfo
  {title} {{Data release for Tests of General Relativity with GWTC-3}},\ }\href
  {https://doi.org/10.5281/zenodo.7007370} {10.5281/zenodo.7007370} (\bibinfo
  {year} {2022})\BibitemShut {NoStop}%
\bibitem [{\citenamefont {Stoica}\ and\ \citenamefont
  {Selen}(2004)}]{Stoica_2004}%
  \BibitemOpen
  \bibfield  {author} {\bibinfo {author} {\bibfnamefont {P.}~\bibnamefont
  {Stoica}}\ and\ \bibinfo {author} {\bibfnamefont {Y.}~\bibnamefont {Selen}},\
  }\bibfield  {title} {\bibinfo {title} {Model-order selection: a review of
  information criterion rules},\ }\href
  {https://doi.org/10.1109/MSP.2004.1311138} {\bibfield  {journal} {\bibinfo
  {journal} {IEEE Signal Processing Magazine}\ }\textbf {\bibinfo {volume}
  {21}},\ \bibinfo {pages} {36} (\bibinfo {year} {2004})}\BibitemShut {NoStop}%
\bibitem [{\citenamefont {Pedregosa}\ \emph {et~al.}(2011)\citenamefont
  {Pedregosa}, \citenamefont {Varoquaux}, \citenamefont {Gramfort},
  \citenamefont {Michel}, \citenamefont {Thirion}, \citenamefont {Grisel},
  \citenamefont {Blondel}, \citenamefont {Prettenhofer}, \citenamefont {Weiss},
  \citenamefont {Dubourg}, \citenamefont {Vanderplas}, \citenamefont {Passos},
  \citenamefont {Cournapeau}, \citenamefont {Brucher}, \citenamefont {Perrot},\
  and\ \citenamefont {Duchesnay}}]{scikit-learn}%
  \BibitemOpen
  \bibfield  {author} {\bibinfo {author} {\bibfnamefont {F.}~\bibnamefont
  {Pedregosa}}, \bibinfo {author} {\bibfnamefont {G.}~\bibnamefont
  {Varoquaux}}, \bibinfo {author} {\bibfnamefont {A.}~\bibnamefont {Gramfort}},
  \bibinfo {author} {\bibfnamefont {V.}~\bibnamefont {Michel}}, \bibinfo
  {author} {\bibfnamefont {B.}~\bibnamefont {Thirion}}, \bibinfo {author}
  {\bibfnamefont {O.}~\bibnamefont {Grisel}}, \bibinfo {author} {\bibfnamefont
  {M.}~\bibnamefont {Blondel}}, \bibinfo {author} {\bibfnamefont
  {P.}~\bibnamefont {Prettenhofer}}, \bibinfo {author} {\bibfnamefont
  {R.}~\bibnamefont {Weiss}}, \bibinfo {author} {\bibfnamefont
  {V.}~\bibnamefont {Dubourg}}, \bibinfo {author} {\bibfnamefont
  {J.}~\bibnamefont {Vanderplas}}, \bibinfo {author} {\bibfnamefont
  {A.}~\bibnamefont {Passos}}, \bibinfo {author} {\bibfnamefont
  {D.}~\bibnamefont {Cournapeau}}, \bibinfo {author} {\bibfnamefont
  {M.}~\bibnamefont {Brucher}}, \bibinfo {author} {\bibfnamefont
  {M.}~\bibnamefont {Perrot}},\ and\ \bibinfo {author} {\bibfnamefont
  {E.}~\bibnamefont {Duchesnay}},\ }\bibfield  {title} {\bibinfo {title}
  {Scikit-learn: Machine learning in {P}ython},\ }\href@noop {} {\bibfield
  {journal} {\bibinfo  {journal} {Journal of Machine Learning Research}\
  }\textbf {\bibinfo {volume} {12}},\ \bibinfo {pages} {2825} (\bibinfo {year}
  {2011})}\BibitemShut {NoStop}%
\bibitem [{\citenamefont {Buitinck}\ \emph {et~al.}(2013)\citenamefont
  {Buitinck}, \citenamefont {Louppe}, \citenamefont {Blondel}, \citenamefont
  {Pedregosa}, \citenamefont {Mueller}, \citenamefont {Grisel}, \citenamefont
  {Niculae}, \citenamefont {Prettenhofer}, \citenamefont {Gramfort},
  \citenamefont {Grobler}, \citenamefont {Layton}, \citenamefont {VanderPlas},
  \citenamefont {Joly}, \citenamefont {Holt},\ and\ \citenamefont
  {Varoquaux}}]{sklearn_api}%
  \BibitemOpen
  \bibfield  {author} {\bibinfo {author} {\bibfnamefont {L.}~\bibnamefont
  {Buitinck}}, \bibinfo {author} {\bibfnamefont {G.}~\bibnamefont {Louppe}},
  \bibinfo {author} {\bibfnamefont {M.}~\bibnamefont {Blondel}}, \bibinfo
  {author} {\bibfnamefont {F.}~\bibnamefont {Pedregosa}}, \bibinfo {author}
  {\bibfnamefont {A.}~\bibnamefont {Mueller}}, \bibinfo {author} {\bibfnamefont
  {O.}~\bibnamefont {Grisel}}, \bibinfo {author} {\bibfnamefont
  {V.}~\bibnamefont {Niculae}}, \bibinfo {author} {\bibfnamefont
  {P.}~\bibnamefont {Prettenhofer}}, \bibinfo {author} {\bibfnamefont
  {A.}~\bibnamefont {Gramfort}}, \bibinfo {author} {\bibfnamefont
  {J.}~\bibnamefont {Grobler}}, \bibinfo {author} {\bibfnamefont
  {R.}~\bibnamefont {Layton}}, \bibinfo {author} {\bibfnamefont
  {J.}~\bibnamefont {VanderPlas}}, \bibinfo {author} {\bibfnamefont
  {A.}~\bibnamefont {Joly}}, \bibinfo {author} {\bibfnamefont {B.}~\bibnamefont
  {Holt}},\ and\ \bibinfo {author} {\bibfnamefont {G.}~\bibnamefont
  {Varoquaux}},\ }\bibfield  {title} {\bibinfo {title} {{API} design for
  machine learning software: experiences from the scikit-learn project},\ }in\
  \href@noop {} {\emph {\bibinfo {booktitle} {ECML PKDD Workshop: Languages for
  Data Mining and Machine Learning}}}\ (\bibinfo {year} {2013})\ pp.\ \bibinfo
  {pages} {108--122}\BibitemShut {NoStop}%
\bibitem [{\citenamefont {Bromiley}(2013)}]{Bromiley:2013}%
  \BibitemOpen
  \bibfield  {author} {\bibinfo {author} {\bibfnamefont {P.~A.}\ \bibnamefont
  {Bromiley}},\ }\bibfield  {title} {\bibinfo {title} {Products and
  convolutions of gaussian probability density functions}\ }(\bibinfo {year}
  {2013})\BibitemShut {NoStop}%
\bibitem [{\citenamefont {Hogg}\ \emph {et~al.}(2020)\citenamefont {Hogg},
  \citenamefont {Price-Whelan},\ and\ \citenamefont {Leistedt}}]{Hogg:2020}%
  \BibitemOpen
  \bibfield  {author} {\bibinfo {author} {\bibfnamefont {D.~W.}\ \bibnamefont
  {Hogg}}, \bibinfo {author} {\bibfnamefont {A.~M.}\ \bibnamefont
  {Price-Whelan}},\ and\ \bibinfo {author} {\bibfnamefont {B.}~\bibnamefont
  {Leistedt}},\ }\href@noop {} {\bibinfo {title} {Data analysis recipes:
  Products of multivariate gaussians in bayesian inferences}} (\bibinfo {year}
  {2020}),\ \Eprint {https://arxiv.org/abs/2005.14199} {arXiv:2005.14199
  [stat.CO]} \BibitemShut {NoStop}%
\end{thebibliography}%

\end{document}